\def\year{2019}\relax
\documentclass[journal]{IEEEtran} 

\usepackage{amsmath,amssymb,amsfonts}
\usepackage{algorithmic}
\usepackage{textcomp}
\usepackage{xcolor}
\def\BibTeX{{\rm B\kern-.05em{\sc i\kern-.025em b}\kern-.08em
    T\kern-.1667em\lower.7ex\hbox{E}\kern-.125emX}}

\usepackage{times}  
\usepackage{helvet}  
\usepackage{courier}  
\usepackage{url}  
\usepackage{graphicx}  
\usepackage[textsize=scriptsize,backgroundcolor=red!10,bordercolor=red!90]{todonotes}

\usepackage{mathtools}
\usepackage{subcaption}
\usepackage{flushend}
\usepackage{balance}
\usepackage{multirow}
\usepackage{hyperref}

\begin{document}
%

\title{Detecting and  Characterizing Extremist Reviewer Groups in Online Product Reviews}

\author{\IEEEauthorblockN{Viresh Gupta, Aayush Aggarwal, Tanmoy Chakraborty\\ Dept. of Computer Science \& Engg., IIIT-Delhi, India}\\
\IEEEauthorblockA{\{viresh16118, aayush16002,  tanmoy\}@iiitd.ac.in}}

        

\maketitle
\begin{abstract}
Online marketplaces often witness opinion spam in the form of reviews. People are often hired to target specific brands for promoting or impeding them by writing highly positive or negative reviews. This often is done collectively in groups. Although some previous studies attempted to identify and analyze such opinion spam groups, little has been explored to spot those groups who target a brand as a whole, instead of just products.

In this paper, we collected reviews from the Amazon product review site and manually labelled a set of 923 candidate reviewer groups. The groups are extracted using frequent itemset mining over brand similarities such that users are clustered together if they have mutually reviewed (products of) a lot of brands. 
We hypothesize that the nature of the reviewer groups is dependent on 8 features specific to a (\emph{group,brand}) pair. We develop a feature-based supervised model to classify candidate groups as extremist entities. We run multiple classifiers for the task of classifying a group based on the reviews written by the users of that group, to determine if the group shows signs of extremity. A 3-layer Perceptron based classifier turns out to be the best classifier.
We further study behaviours of such groups in detail to understand the dynamics of brand-level opinion fraud better. These behaviours include consistency in ratings, review sentiment, verified purchase, review dates and helpful votes received on reviews. Surprisingly, we observe that there are a lot of verified reviewers showing extreme sentiment, which on further investigation leads to ways to circumvent existing mechanisms in place to prevent unofficial incentives on Amazon. 
\end{abstract}

\begin{figure*}
    \begin{subfigure}{0.24\textwidth}
        \centering
        \includegraphics[width=\linewidth]{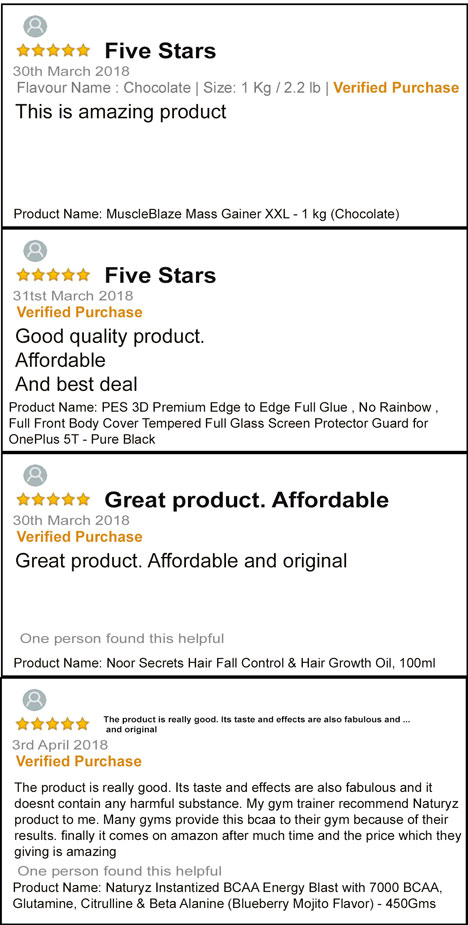}
        \caption{User 1}
        \label{fig:extremist1}
    \end{subfigure}
    \begin{subfigure}{.24\textwidth}
        \centering
        \includegraphics[width=\linewidth]{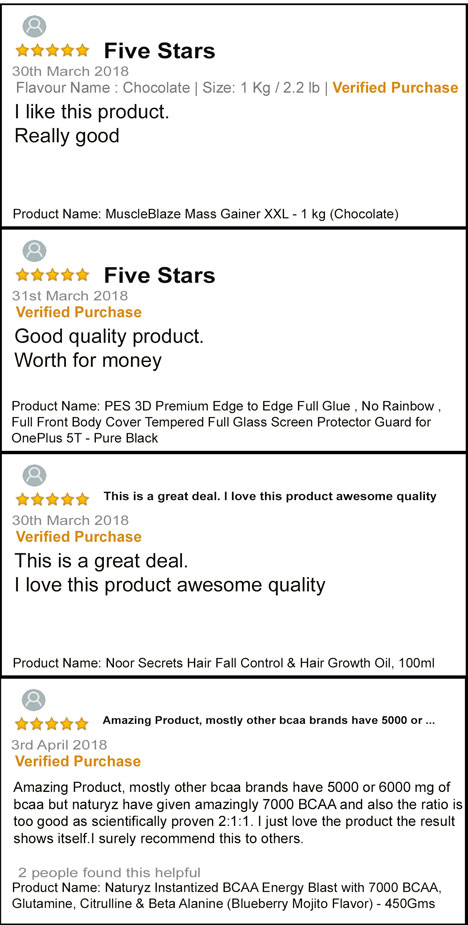}
        \caption{User 2}
        \label{fig:extremist2}
    \end{subfigure}
    \begin{subfigure}{.24\textwidth}
        \centering
        \includegraphics[width=\linewidth]{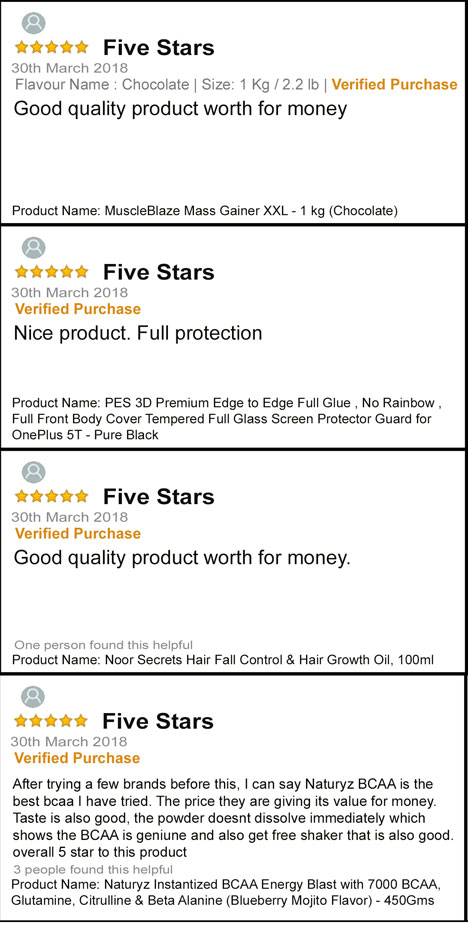}
        \caption{User 3}
        \label{fig:extremist3}
    \end{subfigure}
    \begin{subfigure}{.24\textwidth}
        \centering
        \includegraphics[width=\linewidth]{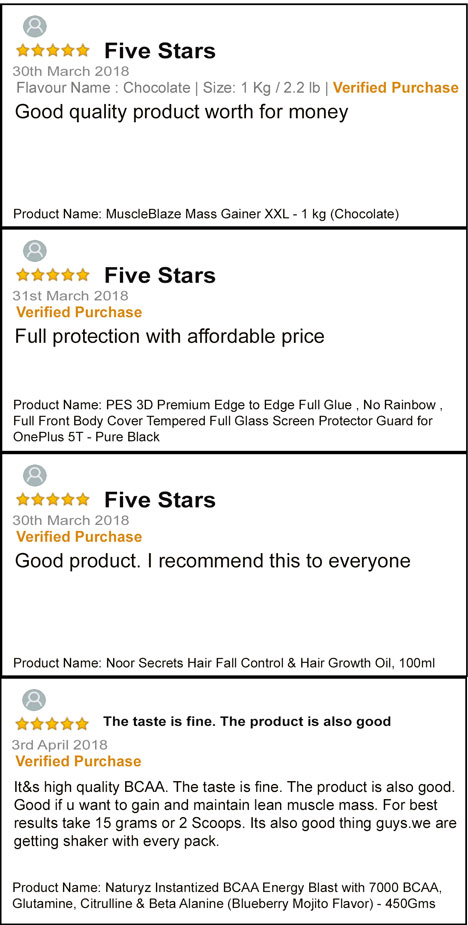}
        \caption{User 4}
        \label{fig:extremist4}
    \end{subfigure}
    \caption{ A sample extremist group.  The reviews are all positive for different brands and also have a verified purchase sign with all of them done on nearly the same dates. For example, in the first row, all reviewers have reviewed the product on the same date, and have given a five-star rating with the review text having extremely similar content.
    }
    \label{fig:example_group}
\end{figure*}

\section{Introduction}
In today's world dominated by online marketplaces, review portals and websites play a crucial role in the buyer's decision for their next purchase. 

\textit{``It’s a virtuous cycle -- the more reviews, the more buys. The more buys, the more reviews. The more buys, the higher your rank in search and the more sales you get,"} says Alice \cite{quoteLink}, the owner of online cosmetic brand Elizabeth Mott.

Undoubtedly, it is highly likely that some people write reviews which are less than truthful to manipulate widespread decision of buyers in their favour.
These people act either individually or in groups. While individual reviewers write such reviews in a matter of frustration or joy, they don't influence the overall opinion on a product to a large extent, but help other buyers by stating their experiences. However, a more compelling case is when multiple individuals form an intricate web, and due to sheer higher number of people reviewing (and certain other techniques, discussed in Section \ref{sec:other_techniques}),
they end up being a major influence on the overall sentiment of the product. \textcolor{black}{The extent of such influence is not just limited to the reviews by opinion spam. Previous work \cite{review_echo_chamber} has shown that 10-15\% reviews are essentially echoing the earliest reviews and thus a misleading early review has an even higher influential potential.}

This is wide-spread opinion spam, and every review website must be aware of this activity and take appropriate measures for identification and/or prevention of this phenomenon. This is a classic example of collective fraud behaviour, where several users are part of a business network and work together to target and influence a particular product. This is a lesser-known phenomenon, and most groups work following certain techniques to not make their collaboration obvious. However, since such groups are economically or otherwise incentivised, and several of these are generally run by a given organization, they have several targets for opinion spam which often share certain common characteristics in their nature of reviews. These characteristics can be exploited to classify them better using a robust and thorough analysis technique.
Amazon India, to prevent opinion spam, has brought about a new policy which limits the number of reviews on a product in a day, as stated in \cite{ReviewLimit}. 

%
In order to still be effective, we claim that certain groups target brands in general, and post extreme reviews across multiple products for a given target brand. This is a higher level of opinion spamming, deliberately writing highly positive or negative reviews for a brand in general in order to promote or demote them in the cut-throat competition of the online marketplace. Studies have been conducted to identify such groups which try to influence a product \cite{GSRank, Lu_annotation_simult, Speagle}; however, groups exhibiting a brand-based opinion spamming is a phenomenon that remains widely unexplored. A detailed discussion is required for these brand-related activities because these practices are against the code of conduct of these review websites, since they negatively skew the brand-based competition, giving innate (dis)advantages to certain brands.

%
Since only the non-verified reviews are limited by the policies\footnote{https://www.amazon.in/gp/help/customer/display.html?nodeId=201930110},
reviewers from these groups can often purchase the product via Amazon in exchange for unofficial discounts (e.g., cash backs) and post verified reviews since they did not receive a discount via amazon's mechanisms (e.g., coupons) (see further discussed in Section \ref{sec:other_techniques}).

Figure \ref{fig:example_group} shows an example of such extremist groups (taken from our annotated dataset as mentioned in Section \ref{sec:dataset}). Four rows correspond to products belonging to four different brands. Four columns represent four different reviewers who, according to our annotation, are part of the same group. Each box represents the review information. This is an example of reviewers showing extreme likeliness towards these products/brands as can be seen from the extreme ratings, similar comments and almost same date \cite{sarthikaijcai}. 
It is clear that this group of reviewers had extreme sentiments towards the brands reviewed, both in terms of the ratings and the review content. 
While this has a large overlap with the classic collusion in which reviewers target a product to bring up its ratings, this kind of extremism in reviews is not in order to promote/demote the ranking of a \textit{product}, but rather to influence the perception of people for a \textit{brand}. 

\textcolor{black}{It is worth noting that such a kind of characterization is different from just combining the groups of people who provide extreme reviews on a product, because while the groups focusing on a product may be extreme in their opinion, they don't necessarily intend to influence the brand image. This, coupled with the fact that the same product may be sold by different sellers, reduces the chance that a product level opinion spamming reviewer group would target products of the same brand. Sellers may not have any inclination towards promoting any particular brand's products; rather, they would prefer to gain a better revenue on all products (may belong to different brands) by their promotional campaigns. Hence such a group is more likely to have a very narrow and specialised source, e.g., the competing brands or the manufacturing brand itself.}

In this paper, we identify and study the behavioural characteristics of extremist reviewer groups. We also build a feature-based classifier based on the brand-specific activities of reviewer groups to identify extremist groups on the Amazon India marketplace. We then further analyze our methodology to unfold behaviours which best signify such activities, and compare and analyze the overall trend of these groups viz-a-viz their behaviours.

The major contributions of this paper are fourfold:
\begin{itemize}
    \item A manually labelled dataset of $923$ reviewer groups which are classified into `extremist' and `moderate' categories.
    \item The first-ever characterization and study of the novel problem of identifying brand-level extremism.
    \item Detailed characterization of extremist reviewer groups.
    \item Design supervised approach to detect extremist reviewer groups.
\end{itemize}

To encourage reproducible research, we have made the codes publicly available at {\color{black}\url{https://github.com/virresh/extremist-reviewers}}.\footnote{We could not release the original Amazon Data due to the copyright issue. Instead, we have released the scraper to collect the data. We have also released the annotated data and feature values of individual groups. }

\textcolor{black}{The paper is organized as follows:
We briefly survey the various studying related to review extremism, applications developed using reviews and fake review detection in Section \ref{sec:background}. Section \ref{sec:dataset} outlines the details of collected dataset and annotation methodology. Section \ref{sec:classification_model} presents the   modeling of features at brand level and  features that relate to brand level extremism in groups along with the experimental setup. Section \ref{sec:results_classification} briefly discusses the results obtained by various classifiers. Sections \ref{sec:feat_imp} and \ref{sec:characterization} characterize extremist users based on the features and expert annotation. Section \ref{sec:other_techniques} discusses the implications of the study. Finally, we conclude in Section \ref{sec:conclusion}.
}


\section{Background and Related Work}\label{sec:background}



We divide the existing literature into two parts: general studies on e-commerce services and detection of fake reviews. 

\subsection{General Studies on E-commerce Reviews}
There have been extensive studies on mining online reviews and classifying them based on user sentiment \cite{Dave:2003:MPG:775152.775226, Pang:2002:TUS:1118693.1118704, 6508366, YE20096527}.
\textcolor{black}{
Reviews have also been extensively used in developing and augmenting recommendation systems \cite{Chelliah:2017:PRE:3109859.3109936, CHEN201344, recommendation_with_review, review_with_collabfiltering} and extracting product features \cite{Popescu:2005:EPF:1220575.1220618, Liu:2005:OOA:1060745.1060797, Hu:2004:MOF:1597148.1597269}. Another study has also shown the utility of product reviews in explaining the recommendations given by a recommendation system \cite{explain_recommendation}. 
Pang et al. \cite{OpinionMiningAndSentimentAnalysis} showed the progression of reviews as an important part of the decision-making process with the advent of web 2.0 and studied them from retrieval perspective. Since it is difficult for the buyer to wade through volumes of reviews, researchers have conducted studies on summarizing reviews based on user sentiment \cite{Hu:2004:MSC:1014052.1014073} and other features \cite{Somprasertsri2010MiningFI, Zhuang:2006:MRM:1183614.1183625, pecar-2018-towards} as well under the umbrella of opinion summarization. 
}
All these studies indicated that product reviews are an invaluable resource for determining the quality of a product. Various marketing studies have also shown that reviews play an important role in maintaining the online reputation of a brand as well \cite{HoDac2013TheEO, boxofficeMarketingStudyWordOfMouth}.

\textcolor{black}{A review usually consists of a star rating which helps to influence a product's overall ratings, but a review becomes even more impactful when people read it. It has been found that people read a review only when it is perceived as helpful by them, which may be through various means -- the helpful upvotes by other consumers, the length of the review, star ratings, readability, etc. \cite{extremity_helpful_review, Yin2014ExploringHC}.  There have been several efforts to understand what makes a review helpful to people \cite{extremity_helpful_review, Kim:2006:AAR:1610075.1610135, Wang2019}. Based on these results, attempts have been made to construct systems that can recommend reviews to users \cite{recommendReviews} by predicting the perceived helpfulness of a review. An important point to note is that all these studies emphasize that review extremity is an important factor for assessing the impact made by the reviews.} 

\textcolor{black}{Extensive work has been done on assessing how review extremity affects users \cite{twosided, extreme_ebay, extreme_book} and the relation between star ratings given by a reviewer and the reviewer's attitude \cite{Krosnick1993}. A moderate review can arise out of indifference or ambivalence \cite{Kaplan_1972, Presser_1980}, and could prove to be useful for buyers seeking in-depth knowledge \cite{twosided}, and an extreme review provides quick validation for experience goods like movies, books, etc. \cite{extreme_ebay, extreme_book}. It has been observed that moderate reviews can affect ``\textit{brand attitude}" \cite{extremity_helpful_review}. All the work indicates extremity in reviews on a product level, which affects the brand as a whole. We make an attempt to understand extremity in reviews directly at the brand level and associate it with groups involved in the same.}


\subsection{Studies on Fake Reviews}
\textcolor{black}{Since reviews are such an impactful resource, it is to be expected that the review space is infested with malpractices as well. There have been various efforts to uncover these practices and understand them in-depth, broadly called as opinion spam.} These studies can be broadly classified into three categories:

\subsubsection{Studies on Reviews}
Jindal and Liu \cite{WSDMJindal} made a pioneering effort to detect fake reviews. They introduced the problem of opinion spam and analyzed online reviews in three varieties - untruthful opinions, seller/brand only reviews (no product involved) and non-reviews using near-duplicate content as an indicator of fake reviews. Other studies dealing with the detection of review-level spam explored linguistic features of text \cite{Ottreviews2}, hand-made rules \cite{onlyReviewSpam2} and combination of review and reviewer features \cite{onlyReviewSpam3}. A probabilistic framework for the same has also been proposed in \cite{MarkovFields}. \textcolor{black}{Ott et al. \cite{Ottreviews2} synthesized fake hotel reviews using Amazon Mechanical Turk, whereas Jindal and Liu  \cite{WSDMJindal} worked on data scraped from Amazon and used content duplicity as ground-truth. Both of them worked with features at a review level. Jindal et al. \cite{onlyReviewSpam2} and Li et al. \cite{onlyReviewSpam3} mentioned the role of brands briefly, but the main focus was on fake reviews rather than extreme reviews.}

\subsubsection{Studies on Reviewers}
Studies involved in detecting reviewer fraud consider rating behaviour \cite{ratingBehaviours}, trust scores based on a relationship graph among reviewers, reviews and stores \cite{onlyReviewer1}. Other studies \cite{onlyReviewer2, MukherjeeFootprinting, burstiness} provided various other methods exploiting behavioural footprints to identify fraud reviewers such as bursts of popularity and use Bayesian approaches. \textcolor{black}{Notably, Wang et al. \cite{onlyReviewer1} introduced the usage of a \textit{review graph} for identifying such spammers. Mukherjee et al. \cite{onlyReviewer2} tried to uncover features used by Yelp filters for abnormal behaviours and revealed that reviewers involved in writing fake reviews showed behavioural features and psychological patterns of overuse of top common words. Mukherjee at al. \cite{MukherjeeFootprinting} and Fei et al. \cite{burstiness} employed strategies like Bayesian modelling of spamicity as latent behaviour and Loopy Belief propagation on modelled Markov Random Fields. However, in all the approaches, the indicators for a fake reviewer, especially when characterizing a reviewer by their rating behaviours \cite{ratingBehaviours}, highly positive or highly negative reviewers were more often a suspect than a moderate rating. This may indicate that extremism in reviews might also be related to opinion spam; however, this front has not been explored yet. }

\subsubsection{Studies on Reviewer Groups} The effect of fraud reviewer groups is more detrimental and subtle than individual
fraud reviewers. 
The issue of manual labelling was addressed by considering a group of reviewers instead of individual reviews. Mukherjee et al. \cite{GSRank,kumar2018rev2} showed that labelling a group of reviewers is considerably easier than labelling individual reviews. Other interesting studies that leverage metadata to characterize different entities in e-commerce sites can be observed in \cite{Speagle, Lu_annotation_simult}, where products, reviews and users are classified simultaneously. Fei et al. \cite{burstiness} and Kakhki et al. \cite{buyaccountbatches} showed that synchronicity is an important group behaviour; Xu and Zhang \cite{Xuhcbm} further used this signal as a temporal indicator alongside combining several other measures and proposed a completely unsupervised model for detecting group collusion. \textcolor{black}{Several graph-based approaches have also shown the potential to detect both spam reviewers and spam reviews simultaneously \cite{Lu_annotation_simult}. Wang et al. \cite{GGSpam} and Dhawan et al.  \cite{sarthikaijcai} extended the reviewer graph to detect collusive users, i.e., a temporary group of users that work in unison to spam. Again there has been no study about any phenomenon of extremism at a group level, especially with respect to a brand, since extremism ultimately affects ``\textit{brand attitudes}''.}

\subsection{How Our Work is Different from Others?}
Our work is different from the existing studies in the following ways: 
\begin{itemize}
    \item We introduce the problem at a brand level which was not considered in any of the previous studies.
    \item Unlike other studies which majorly focus on fake review/reviewer detection, we here focus on extremist reviewer detection, which may not be fake. Moreover, we attempt to identify `groups' instead of detecting `individual user'.  
    \item We investigate the effect of Amazon's 2016 changes in reviewing policy and the review scenario post policy changes.
\end{itemize}

\begin{table}[!t]
\centering
\caption{\label{DatasetDescription}Description of the collected and annotated dataset.}
\scalebox{0.9}{
\begin{tabular}{ l|l|l|l|l }
 \hline
 {\bf Category} & {\bf \# Reviews} & {\bf\# Reviewers} & {\bf \# Brands} & {\bf \# Products}\\
 \hline
 All   &   17,24,656 & 10,77,027 & 30,486  & 1,88,298 \\
 \hline
 Filtered by & & & & \\
 reviewer history & 3,41,081 & 10,379 & 28,437 & 1,81,596 \\
 \hline
 Annotated set & 45,892 & 366 & 8,005 & 33,892 \\ 
 \hline
\end{tabular}}
\vspace{-5mm}
\end{table}

\section{Dataset}\label{sec:dataset}

To the best of our knowledge, no dataset of consumer reviews (on Amazon) that consists of brand information exists so far. Thus, we attempted to create the first dataset of its kind by crawling reviews from \url{Amazon.in}, the Indian counterpart of the e-commerce giant. In this dataset, along with the regular metadata, we also obtained the brand on which a review was posted. Other metadata per-review include reviewer id, product id (ASIN), brand, rating, review text, date, and the number of helpful votes a review has received.

We also collected review history of users from their Amazon profiles. Since scraping profile data is considerably slow (due to bot authentications encountered by our scraper on the customer's Amazon profile) compared to product reviews, we collected information for only those users who reviewed at least two products of the same brand. This information would allow us to study the characteristics of the individual reviewer we get and form a characterisation of the reviewers belonging to extreme categories.

The data was crawled during August - October 2018. The dataset containing review history of reviewers consists of $341,081$ reviews from $103,79$ unique reviewers, spanning across $181,596$ products from $28,437$ brands. Table \ref{DatasetDescription} shows the statistics of the complete dataset.

Since the dataset contains brand information associated with each review, these reviews can be used to identify groups of reviewers targeting a particular brand. This problem can be seen as an extension of the existing data mining techniques such as Frequent Itemset Mining (FIM) by considering itemset as a sequence of reviewer IDs, where each transaction contains the IDs of reviewers who have reviewed several products of the same brand.

    
    

\subsection{Preprocessing the Dataset}
On Amazon, usually, a brand does not directly sell its products. There are sellers that post their products and post them up for sale. Thus, the scraped data with brand names had an inherent issue. Different sellers might not give the exact same brand name to products they list. Thus, identifying which products belong to the same group needs some preprocessing since the brand name may have subtle differences such as different case (e.g., Whirlpool vs WhirlPool) or different acronyms (e.g., LG vs L.G. Electronics).

Therefore, we used case-folding followed by careful stemming to match products belonging to the same brand. Another issue was anonymous reviews and deleted products. Since we scraped reviews from user profiles, it is possible that some products would have got deleted. In this case, we did not have any brand information about this product, but the review existed on the user's profile. To deal with these cases, we removed all reviews which did not have a brand associated with them. Since every reviewer, including the anonymous reviewers, is given a unique reviewer id, we did not remove any anonymous reviews.

\subsection{Detecting Candidate Groups}
We use FIM  based on the recursive elimination algorithm \cite{relim} from SPMF library \cite{spmf}, for identifying candidate groups keeping the minimum support count to 15. This would give us a list of groups of reviewers, where all members of a resultant group share at least 15 brands for which they have written at least one review.

However, since members of a given group satisfy the minimum support count criteria of 15, the members of all of its sub-groups also satisfy the same criteria, the algorithm also reports all the sub-groups for a given resultant group.
Therefore, for each group obtained, we prune our result to drop all its sub-groups and keep only the maximal group.\\

\noindent
\textcolor{black}{
\textbf{Selection of minimum support count}\\
Since (group, brand) pairs were to be labeled manually, we wanted to have a decently diverse yet sufficiently large candidate group set. The total number of transactions for FIM to act on (one transaction for every brand) was around $30.4k$. The minimum number of common brands (or minimum support for frequent itemset mining) affects the size of candidate groups drastically as shown in Figure \ref{fig:msp_selection} -- one can notice that beyond the 15 brand mark, the number of candidate groups decreases rapidly. 
At 10 brands ($\approx$ 0.03\%), we have $5568$ unique candidate groups, whereas at 20 brands ($\approx$ 0.06\%) we have $2817$ unique candidate groups. Note that these are unique candidate groups; therefore, during labeling they would be paired with each of the brands that the group has in common. This in turn  increases the amount of (group, brand) pairs considerably. Thus, we choose 15 brands ($\approx$ 0.05\%) as the minimum support count which results in $3746$ unique groups.
\begin{figure}[h]
    \centering
    \includegraphics[width=0.4\textwidth]{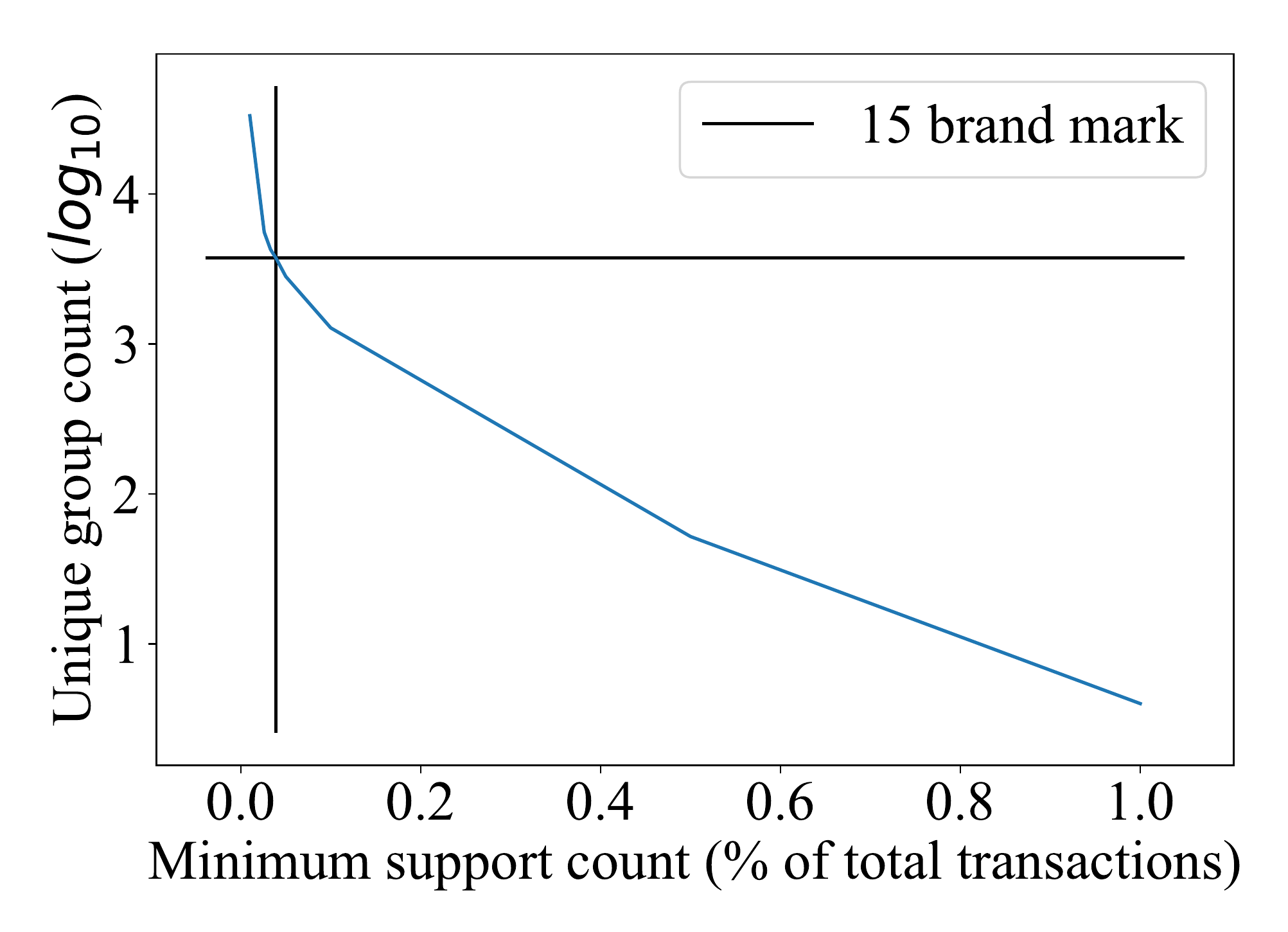}
    \caption{{\color{black}Change in the number of unique groups with the minimum support count. Beyond the 15 brand mark (i.e., $\approx$ 0.05\% of transactions) we can observe the rapid decrease in unique potential candidate groups; whereas before this mark, we have extremely large number of candidate groups. }}
    \label{fig:msp_selection}
\end{figure}
}


\subsection{Preliminary Observation}
\begin{figure}
    \includegraphics[width=\linewidth]{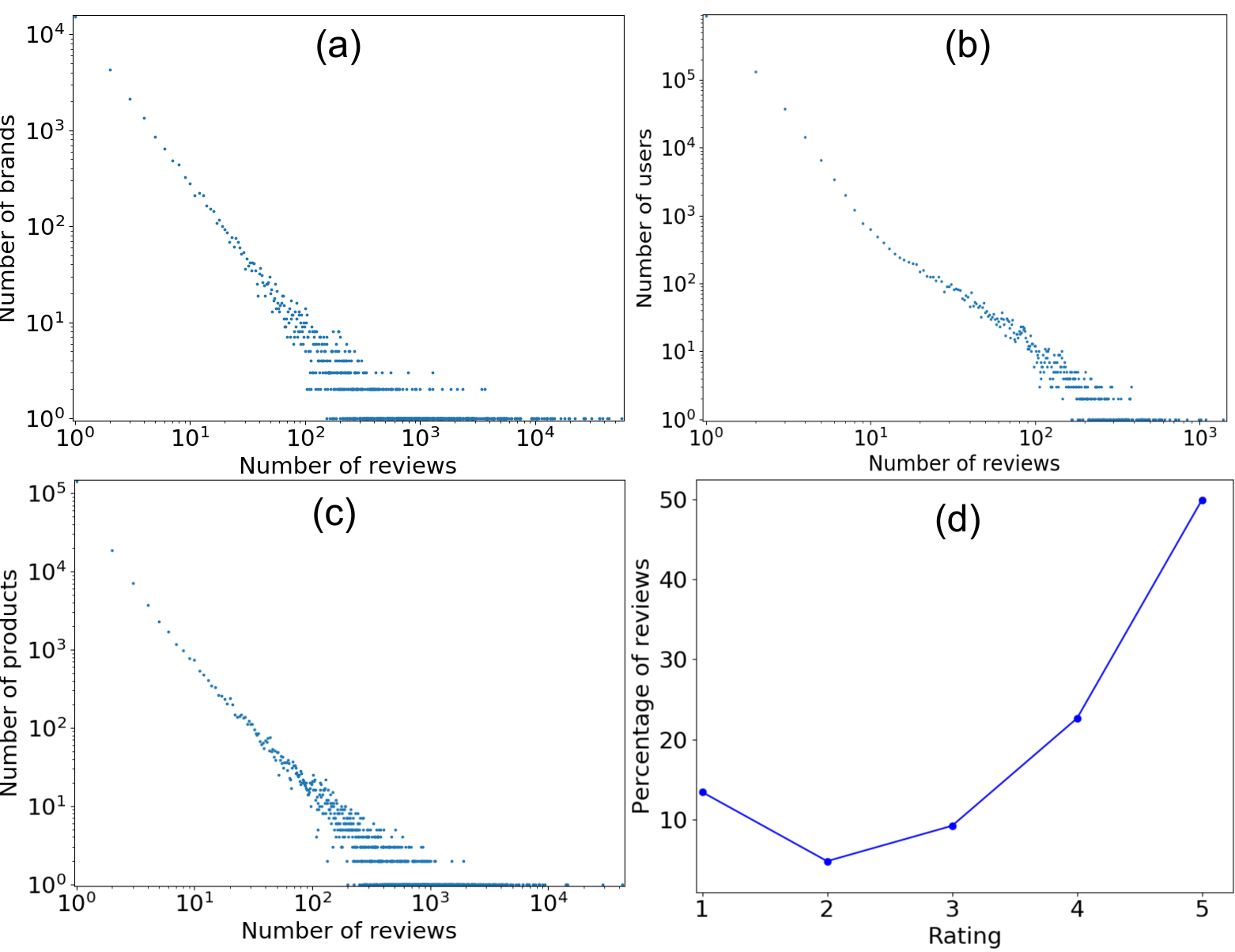}
    \caption{
         Preliminary analysis. Variation of reviews with (a) number of brands, (b) number of reviewers and (b) number of products. We observe a power law distribution in all cases. Fig. (d) shows the percentage of reviews having a given rating.  Also as a general trend, the number of 1- and 2-star reviews is less than 4- or 5-star reviews.}
    \label{fig:powerlaw}
    \vspace{-5mm}
\end{figure}

We perform an analysis of our data to understand the relationships between brand and reviews. To this end, we draw log-log plots of the number of reviews against brands, products and reviewers (c.f. Figure \ref{fig:powerlaw}). We observe a power-law  distribution for not only products and reviewers (as reported in \cite{WSDMJindal}), but also for brands (Figure \ref{fig:powerlaw}(a), \ref{fig:powerlaw}(b) and \ref{fig:powerlaw}(c)). This suggests the possibility that similar methods for detecting extremism at brand level might work. Also, a distribution of the ratings given to reviews is shown in Figure \ref{fig:powerlaw}(d). We observe that the number of reviews with a low rating (1- and 2-star) is comparatively less than that of high rating (4- and 5-star).

\subsection{Human Annotation}
The technique used in previous studies for labelling fake reviews/reviewers was re-crawling to detect the deleted reviews \cite{clustering} and employing human judges for manual labelling \cite{GSRank, liang_annotation1, ratingBehaviours, Lu_annotation_simult, Ottreviews2, Xie_annotation}.

Some crucial insights while manually annotating reviewer groups are mentioned in \cite{annotation}, such as review length, similarity, excessive use of capitalization, brand pin-pointing, personal relationship with brands, etc. The labelling was done keeping in mind these insights regarding the ratings, review title, review text, date posted, the number of helpful votes received, and also in some cases the product and reviewer ids. We leverage this information to perform manual labelling. These have been summarised in Table \ref{annotation_indicators}.

\begin{table}
\caption{Factors kept in mind by the annotators while annotating the dataset.}
\label{annotation_indicators}
\centering
\begin{tabular}{|c|c|} 
\hline
\textbf{Criterion}                           & \textbf{Information focused on}  \\ 
\hline
Brand mentions                               & Individual review                \\ 
\hline
Excessive use of superlatives in text        & Individual review                \\ 
\hline
Excessively positive or negative sentiment   & Individual review                \\ 
\hline
Similarity in product description and review & Individual review                \\ 
\hline
Rating variety                               & Reviewer profile                 \\ 
\hline
Number of reviews written                    & Reviewer profile                 \\ 
\hline
Rating deviation from the average rating     & Reviewer profile                 \\ 
\hline
Reviewer's active duration                   & Reviewer profile                 \\ 
\hline
Similarity in reviews among group members  & Group                            \\ 
\hline
Time of reviews of group members             & Group                            \\ 
\hline
Rank of reviewers the group has              & Group                            \\
\hline
\end{tabular}
\end{table}

Once the pruning is completed by recursive elimination \cite{relim}, we made \emph{(group,brand)} tuples based on what brands the group has reviewed. This gave us around 180,000 pairs. Since
there were too many candidate groups, our annotators
could only manage to label
 923 candidate  \emph{(group,brand)} pairs as either 0 or 1, based on if they are genuine/moderate or extreme (brand critical or brand positive). 469 and 454 groups were marked as extreme and moderate groups, respectively. \textcolor{black}{We employed three annotators who were within the age group of 20-35 years and were experts in e-commerce service policies. Out of 1000 reviews labelled, there was disagreement by the three annotators in around 77 groups. Therefore, we considered only 923 groups which all the three annotators agreed upon. The inter-annotator agreement is $\kappa=0.86$ (Cohen's $\kappa$).}
We then use this as a ground-truth to draw insights about extremist user groups, train classifiers to detect extremist groups, and identify the most important features that such extremist groups exhibit.

One may argue that the annotated dataset is small in size. We would like to emphasize that such annotation is extremely challenging and time-consuming since the annotators need to go through each group separately and analyze the review patterns of all the constituent members of the group.  In our case, the entire annotation took two and a half months to finish. As suggested by \cite{mukherjee2012spotting}, We did not use Amazon Mechanical Turk (MTurk) for this
annotation task because MTurk is normally used to perform simple tasks which require human judgments. However, our task is
highly challenging, and the annotators required access
to our data. Also, we needed annotators who had proper domain knowledge on
the product review domain. Thus, we believe that MTurk was not suitable in our case. Due to the same problem, Mukherjee et al.  \cite{mukherjee2012spotting} also did not use MTurk for the annotation of fake reviewer groups. 

\textcolor{black}{
\subsection{Product-level vs. Brand-level Analysis}
Since no work at product-level deals with extremism, we use spamicity, an unsupervised metric computed by FraudEagle \cite{fraudeagle}, as a substitute to product-level extremism which helps us to (i) compute scores for all reviewers, and (ii) circumvent the problem of labels. Also in this experiment, we only consider the (group, brand) pairs that are labelled as extremist. We compute the spamicity of all reviewers in our database using complete data and then consider the following distributions of spamicity score:
\begin{enumerate}
    \item Distribution of spamicity scores of all reviewers in a (group, brand) pair,
    \item Distribution of spamicity scores of all reviewers for the corresponding brand.
\end{enumerate}
Since the cardinality of the above two distributions might not be same, we take the size of the smaller set (i.e., the number of reviewers in (group, brand) pair) as the maximum rank till which the divergence of the two distributions is computed. 
Figure \ref{fig:brandvprod}(a) shows the Gaussian kernel density estimate of the KL-divergence scores between the two distributions for all extremist (group, brand) pairs. The spamicity scores lie in the range of 0-1, whereas the mean ($\pm$ standard deviation) of the density is $1.48 (\pm0.83)$, which is quite high, indicating the inherent difference in the two distributions.
Figure \ref{fig:brandvprod}(b) shows the absolute spamicity scores plotted according to corresponding ranks. All (group, brand) rankings were flattened out, thus the highest spamicity score amongst all reviewers in (group, brand) pair is compared to the highest spamicity score amongst all reviewers for that brand; and the lowest spamicity score amongst all reviewers in (group, brand) pair is plotted against the same ranked score amongst all reviewers for that brand.
From both of these experiments, we conclude that there is a  significant difference amongst brand-level extremist reviewers to product-level spammers (the spamicity scores do not reflect brand-level extremism), reinforcing the utility of our study.
\begin{figure}
    \centering
    \includegraphics[width=0.9\columnwidth]{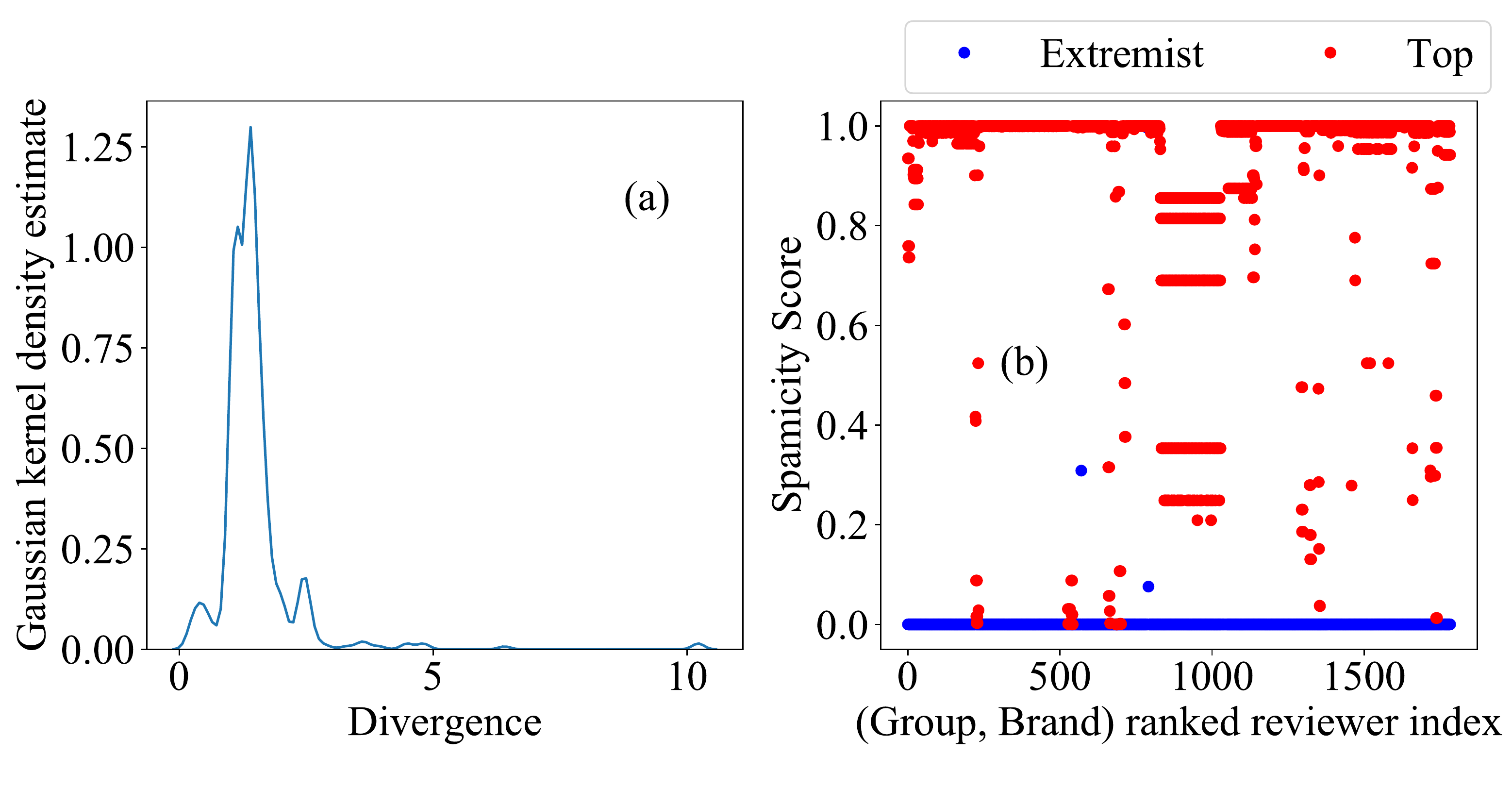}
    \caption{{\color{black}(a) The estimated density of KL divergence in scores between top brand spammers vs. reviewers in extremist (group, brand) pairs for all extremist labeled pairs. The divergence in the spamicity scores for extremist reviewers vs. top spammers is high (peaking around 1.48). (b) The absolute values of these spamicity scores of reviewers. Again the spamicity scores of extremist reviewers (blue) is in extreme contrast with top brand spammers (red), indicating that spamicity at product-level is unable to map extremism at brand level.}}
    \label{fig:brandvprod}
\end{figure}
}

\section{Classification Model} \label{sec:classification_model}
We follow a feature-based classification technique to determine if a candidate group is an extremist group w.r.t to a given brand. For classification, we use state-of-the-art classifiers such as SVM, RandomForest, Logistic Regression, Decision Tree, Gaussian Naive Bayes, K Nearest Neighbors (KNN), Stochastic Gradient Descent (SGD), 3-layer Perceptron (MLP) and XGBoost. \textcolor{black}{We additionally test with an additional layer in the MLP based architecture which consists of two fully connected hidden layers followed by a dense layer with a softmax activation as output layer, forming a 4-Layer Perceptron. Note that this 4-Layer Perceptron is just to test the effect of increased depth.}


\subsection{Features Used for Classification}
Previous studies proposed useful features for detecting group spam. For example, Mukherjee et al. \cite{GSRank} proposed eight features that are highly useful for detecting group spam. We extend these features into brand-level and identify groups of extremist users. The features we used are as follows:

\begin{enumerate}
    \item \textbf{Average Rating:} It captures the average rating given by the group $G$ to a certain brand $B$. We aggregate the reviews given by group members to products of the given brand and take the mean of these ratings. We hypothesize that an extreme group may give an average rating value at the extremes, i.e., closer to 5-star or 1-star.
    \[ avg\_rating(G,B) = \underset{m \in G, p \in B}{avg} rating(m, p) \] where $rating(m,p)$ is the rating given by member $m$ to product $p$.
    
    \item \textbf{Average Upvotes:} It captures how many upvotes the given group receives with respect to the given brand. This is the mean of upvotes taken across reviews posted by the group members, for products belonging to the given brand. 
    \[ avg\_upvotes(G,B) = \underset{m \in G, p \in B}{avg} upvote(m, p) \]
    where $upvote(m,p)$ is the number of upvotes of the reviews for product $p$ posted by reviewer $m$. 
    
    \item \textbf{Average Sentiment:} We analyze the review text for the given \emph{(group,brand)} pair, and find out the average sentiment of these reviews. We use SentiWordNet 3.0 \cite{sentiwordnet3} for the sentiment analyser, which returns an overall sentiment for the review text between -1 to 1. An extremist group may write reviews having an overall sentiment of being highly positive (+1), or highly negative (-1) towards a particular brand.
    \[ avg\_sentiment(G,B) = \underset{m \in G, p \in B}{avg} sentiment(m, p) \]
    where $sentiment(m, p)$ indicates the sentiment of the review on product $p$ posted by reviewer  $m$.
    
    \item \textbf{Group Time Window (GT):}  It indicates the difference between the latest review posted by the group and the earliest review posted by the group for the given brand. A lower value of GT may suggest that the group is closely linked together and indulges in spamming the reviews together.
\[    
{\tiny
   GT(G, B) =\\ 
    \begin{cases}
        \hfil 0                         &  \text{ if $L(G, B)-F(G, B) > \tau $}\\
        1 - \frac{L(G, B) - F(B, B)}{\tau}  &  \text{Otherwise}
    \end{cases}
 }   \]
    
    where $L(G,B)$ and $F(G,B)$ are the last and the first dates of any review posted by any member of group $G$ on any product of brand $B$, respectively. Empirically, $\tau=0.28$ was found to produce the best result.
    
    \item \textbf{Review Count:} This feature captures how many reviews are written by the group for the particular brand. An extremist group is more likely to write more reviews collectively than other users.
    \[ RC(G, B) =  \underset{m \in G}{\sum} |reviews(m, B)| \]
    where $|reviews(m, B)|$ is the number of reviews written by reviewer $m$ on all products belonging to brand $B$.
    
    \item \textbf{Rating Deviation:} It captures how much does the \emph{(group, brand)} deviate from the mean rating. The reviewers of an extremist group are expected to possess a lesser deviation since they must write highly coherent opinions for the brand.
    \[
        \sigma(G, B) = \frac{| \langle r_{G, B} \rangle -  \langle \bar{r}_{G, B} \rangle |}{4}
    \]
    where $\langle r_{G,B} \rangle$ and $\langle \bar{r}_{G,B} \rangle$ represent the average rating given by group $G$ on products belonging to brand $B$, and average rating given by anyone not belonging to group $B$ on products of brand $B$. The deviation is normalized by $4$ since the  rating ranges from 1 to 5; the maximum rating deviation can be 4.
    
    \item \textbf{Early Time Window (ET):} It measures the time gap since the product spawned on the marketplace, and the last review posted on for it by the group. The mean value is taken across all the products for the brand.
    
    {\small
    \[
        ET(G, B) = \underset{p \in B}{avg} \ \  TW ( G, p ) 
    \]
    \[
        TW(G, p) = 
        \begin{cases}
            \hfil 0             &   \text{if $L(G, p) - \alpha (p) > \beta$}\\
            1 - \frac{L(G, p) - \alpha (p)}{\beta} &     \text{Otherwise}
        \end{cases}
    \]
    }
    where $L(G,p)$ is the latest review date for a review done by any member of group $G$ on a product $p$, and $\alpha (p)$ is the earliest review date for product $p$.  Empirically, $\beta=0.28$ was found to produce the best result.
    
    \item \textbf{Verified Purchase:} A review where the product was actually bought by the reviewer holds more credibility than the opposing case.
    This feature determines the fraction of reviews posted by the group for the brand corresponding to amazon-verified-purchase reviews.
    
    \[
    VP(G, B) = \frac{|verified(G, B)|}{RC(G,B)}
    \]
    where $|verified(G,B)|$ is the number of verified reviews done by any member in $G$ for a product belonging to brand $B$, and $RC(G,B)$ is the review count as mentioned earlier.
    
    
    
\end{enumerate}

\begin{table*}[h]\centering
\caption{\label{Accuracy} Performance of various classifiers. The task is of binary classification and thus the values of macro and micro average are both extremely close differing only because of averaging over folds.}
\begin{tabular}{ |p{3cm}||p{1.4cm}|p{1.2cm}|p{1.2cm}|p{1.6cm}||p{1.4cm}|p{1.2cm}|p{1.2cm}|p{1.6cm}| }
\hline
\multirow{2}{*}{\bf Classifier} & \multicolumn{4}{c||}{\bf Micro Average} & \multicolumn{4}{c|}{\bf Macro Average}  \\
\cline{2-9}
                            & {\bf Precision} & {\bf Recall} & {\bf F1} & {\bf ROC-AUC} & {\bf Precision} & {\bf Recall} & {\bf F1} & {\bf ROC-AUC} \\
\hline
SVM                         & 0.930     & 0.930      & 0.930  & 0.930       & 0.940         & 0.927      & 0.933  & 0.927       \\
Logistic Regression         & 0.803     & 0.803      & 0.803  & 0.802       & 0.803         & 0.802      & 0.802  & 0.802       \\
Random Forest               & 0.780     & 0.780      & 0.780  & 0.782       & 0.784         & 0.782      & 0.783  & 0.782       \\
Decision Tree               & 0.699     & 0.699      & 0.699  & 0.701       &0.704          &0.701       & 0.703  & 0.701     \\
Gaussain Naive Bayes        & 0.563     & 0.563      & 0.563  & 0.576       & 0.737         & 0.576      & 0.647  & 0.576       \\
SGD                         & 0.824    & 0.824    & 0.824    & 0.829    & 0.866    & 0.829    & 0.847    & 0.829       \\
KNN                         & 0.814    & 0.814    & 0.814    & 0.816    & 0.818    & 0.816    & 0.817    & 0.816       \\
3-Layer Perceptron                 & \textbf{0.982}    & \textbf{0.982}    & \textbf{0.982}    & \textbf{0.982}    & \textbf{0.982}    & \textbf{0.982}    & \textbf{0.982}    & \textbf{0.982}       \\
4-Layer Perceptron     & 0.914    & 0.914    & 0.914    & 0.917    & 0.925    & 0.917    & 0.921    & 0.917        \\
XGBoost                     & 0.710     & 0.710     & 0.710  & 0.713       & 0.722      & 0.713     & 0.718     & 0.713  \\
\hline
\end{tabular}
\end{table*}

\begin{table*}[h]\centering
\caption{\label{Accuracy_drop} Accuracy of the 3-layer Perceptron after dropping each feature in isolation.}
\begin{tabular}{ |p{3cm}||p{1.4cm}|p{1.2cm}|p{1.2cm}|p{1.6cm}||p{1.4cm}|p{1.2cm}|p{1.2cm}|p{1.6cm}| }
\hline
\multirow{2}{*}{\bf Feature} & \multicolumn{4}{c||}{\bf Micro Average} & \multicolumn{4}{c|}{\bf Macro Average}  \\
\cline{2-9}
                            & {\bf Precision} & {\bf Recall} & \bf{F1} & {\bf ROC-AUC} & {\bf Precision} & {\bf Recall} & {\bf F1} & {\bf ROC-AUC} \\
\hline
Review count               & \textbf{0.640}    & \textbf{0.640}    & \textbf{0.640}    & \textbf{0.641}    & \textbf{0.647}    & \textbf{0.641}    & \textbf{0.644}    & \textbf{0.641}       \\
Rating deviation           & 0.915    & 0.915    & 0.915    & 0.914    & 0.928    & 0.914    & 0.921    & 0.914    \\
Avg. upvotes               & 0.955    & 0.955    & 0.955    & 0.956    & 0.957    & 0.956    & 0.956    & 0.956    \\
Group time window          & 0.962    & 0.962    & 0.962    & 0.962    & 0.963    & 0.962    & 0.963    & 0.962    \\
Verified purchase          & 0.966    & 0.966    & 0.966    & 0.967    & 0.967    & 0.967    & 0.967    & 0.967    \\
Avg. sentiment             & 0.969    & 0.969    & 0.969    & 0.969    & 0.969    & 0.969    & 0.969    & 0.969    \\
Early time window                & 0.971    & 0.971    & 0.971    & 0.971    & 0.972    & 0.971    & 0.972    & 0.971    \\
Avg. rating          & 0.973    & 0.973    & 0.973    & 0.973    & 0.973    & 0.973    & 0.973    & 0.973    \\
\hline
\end{tabular}
\end{table*}

\begin{table}[h]
\centering
\caption{\label{Feature Weights} Weights given to features by Random Forest Classifier.}
\begin{tabular}{ c|c }
 \hline
 {\bf Feature} & {\bf Weight} \\\hline
 Review Count   &\textbf{0.456}\\
 Early time window    &0.154\\
 Verified purchase    &0.133\\
 Avg. rating   &   0.105\\
 Avg. sentiment &0.055\\
 Rating deviation    &0.051\\
 Avg. upvotes   &   0.044\\
 Group Time window    &0.001\\
 \hline
\end{tabular}
\vspace{-5mm}
\end{table}

\begin{figure*}
    \centering

    \begin{subfigure}{0.23\textwidth}
        \includegraphics[width=\linewidth]{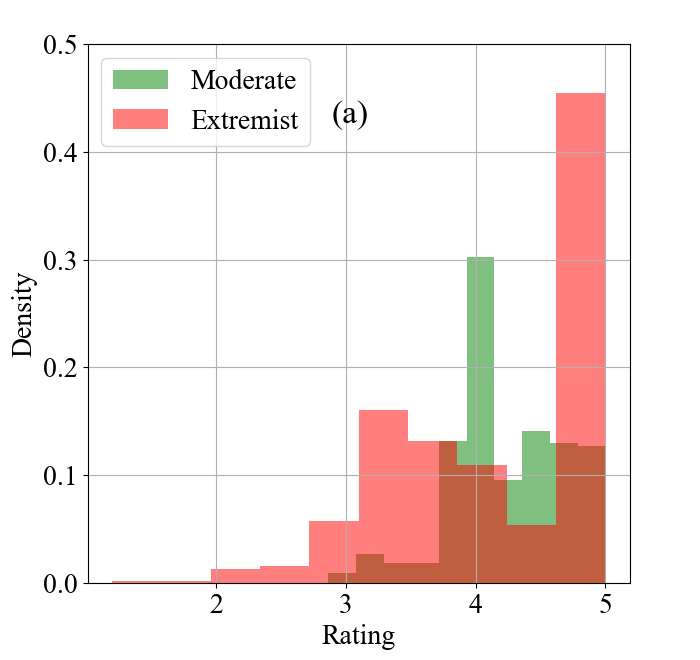}
    \end{subfigure}
    \begin{subfigure}{0.23\textwidth}
        \includegraphics[width=\linewidth]{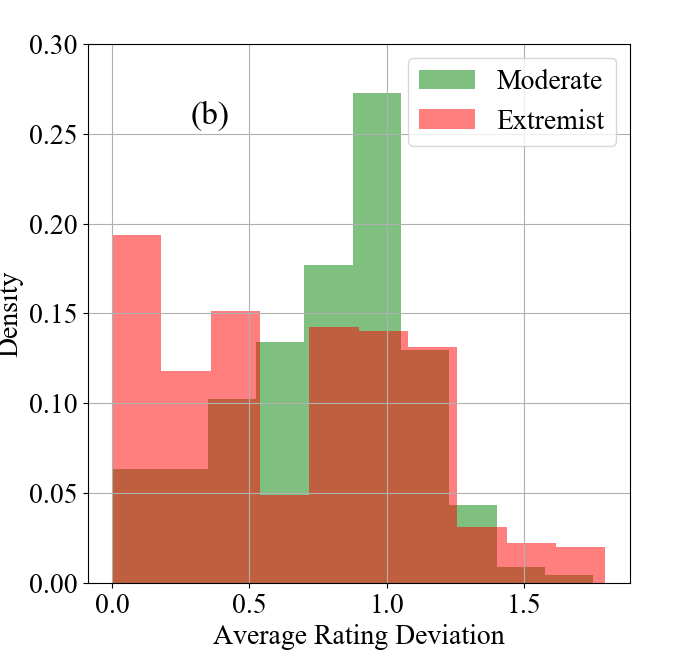}
    \end{subfigure}
    \begin{subfigure}{0.23\textwidth}
        \includegraphics[width=\linewidth]{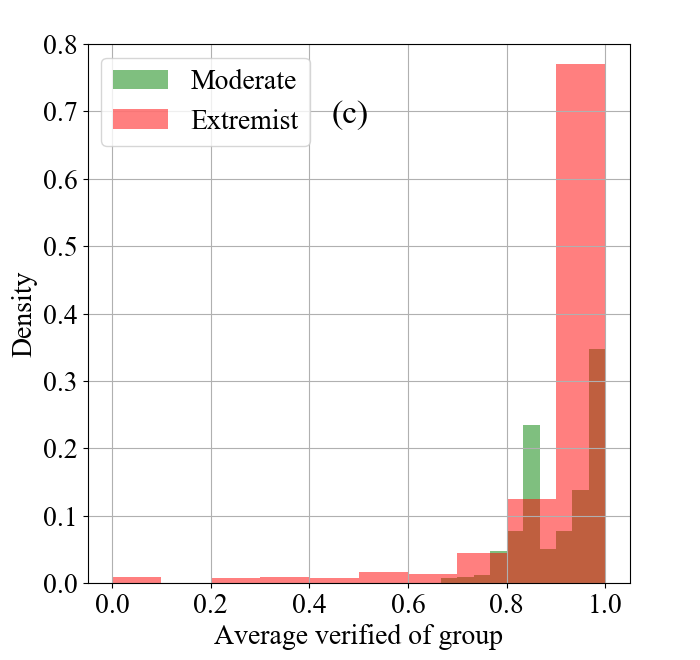}
    \end{subfigure}
    \begin{subfigure}{0.23\textwidth}
        \includegraphics[width=\linewidth]{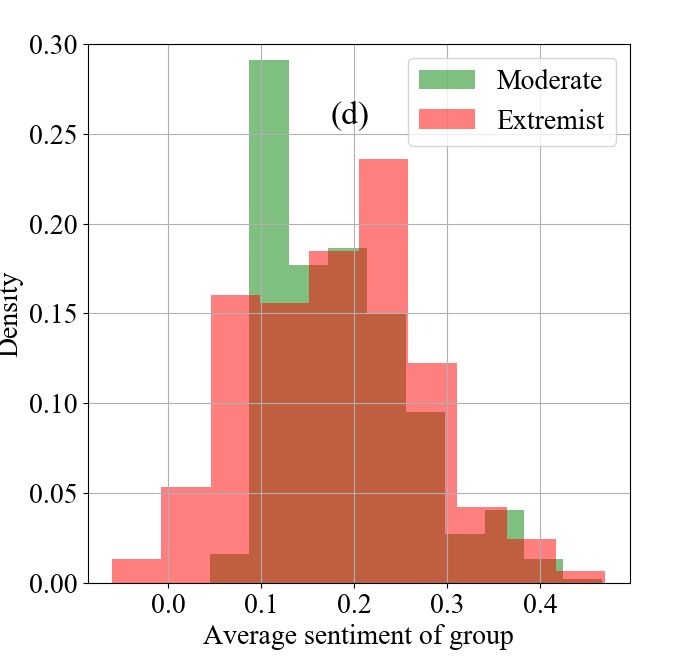}
    \end{subfigure}
    \begin{subfigure}{0.23\textwidth}
        \includegraphics[width=\linewidth]{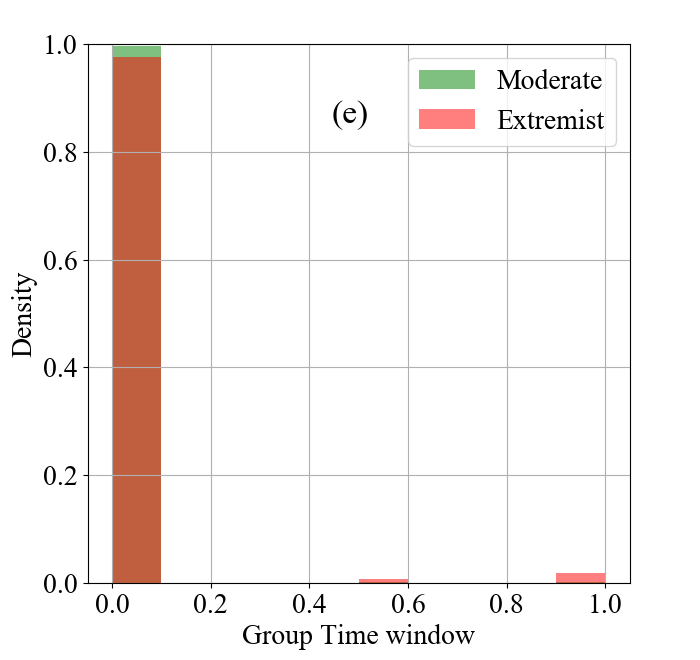}
    \end{subfigure}
    \begin{subfigure}{0.23\textwidth}
        \includegraphics[width=\linewidth]{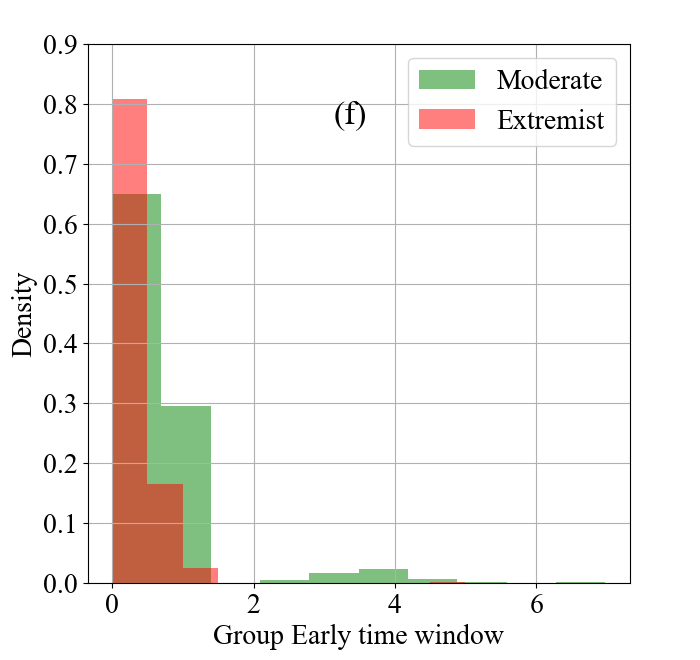}
    \end{subfigure}
    \begin{subfigure}{0.23\textwidth}
        \includegraphics[width=\linewidth]{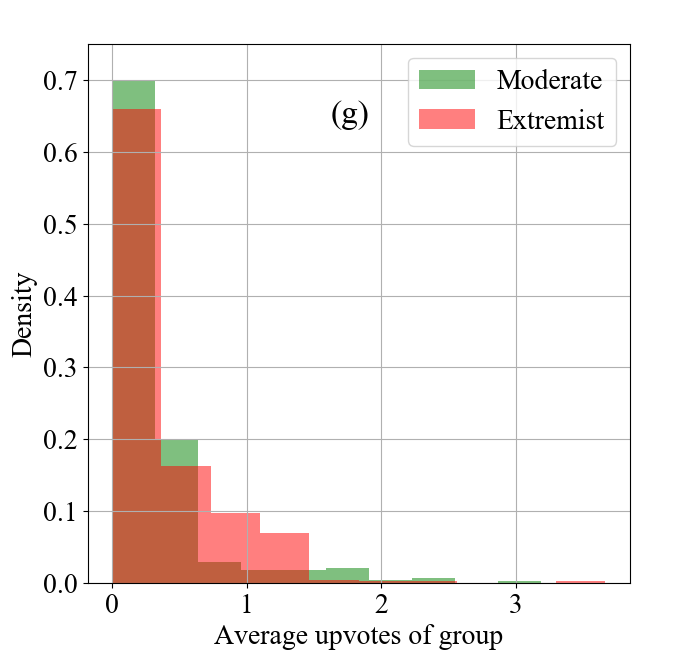}
    \end{subfigure}
    \begin{subfigure}{0.23\textwidth}
        \includegraphics[width=\linewidth]{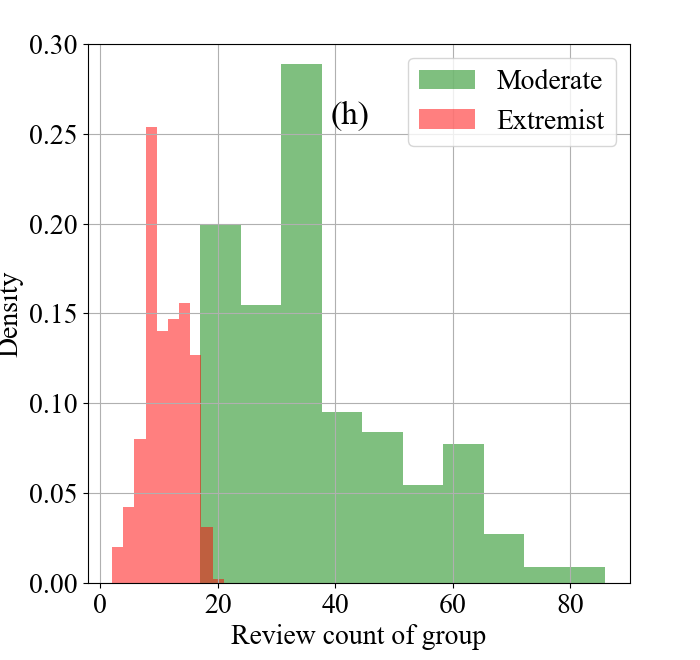}       
    \end{subfigure}
    
    \caption{ (Color online) Comparison of extremist and moderate groups based on different features used in the classification.}
    \label{fig:comparions}
\end{figure*}

\subsection{Experimental Setup}
 We utilize standard grid search for hyperparameter optimization (as suggested in \cite{8793010}). For better reproducibility, we also report the parameters used for each classifier: MLP -- logistic function as activation function, learning rate of $1\text{e-}5$, $2$ hidden layers each with $100$ neurons; KNN with $K=5$; CART with Gini gain criteria for Decision Tree and Random Forest; for SGD learning rate of $1\text{e-}5$, modified Huber as loss function, maximum iteration of $200$, and  L$2$ norm as penalty function, XGBoost with maximum tree depth of 4, learning rate of 0.1 and total of 100 estimators. The results are reported after $10$-fold cross-validation with suitable hyper-parameter optimization.

To report the accuracy of the classifiers, we use standard evaluation metrics, i.e.,  precision, recall, F1-score and Area under the ROC curve (ROC-AUC), with both micro and macro averaging. 

\section{Classification Results}\label{sec:results_classification}

 Table \ref{Accuracy} summarizes the classification accuracy. We observe that all the methods performed decently in predicting the label of the given candidate \emph{(group, brand)} pair. However, the performance of neural network-based methods (3 and 4 Layer Perceptron) is superior compared to the rest of the classifiers. This may be due to the data being accurately mapped by deep learning methods.
 3-Layer Perceptron seems to be the best method with an accuracy of $~0.98$ across all the metrics used to evaluate the performance. Decision Tree seems to be the worst among all the classifiers since it is extremely sensitive to small variations which are prevalent in our dataset. Nevertheless, we conduct the rest of the experiment with 3-Layer Perceptron as the default classifier.

\section{Feature Importance}\label{sec:feat_imp}
To estimate and interpret how much each feature is indicative of extremist behaviour, i.e. how much a feature governs the overall extremity of a group, we use the optimal feature weight distribution of our Random Forest classifier (Table \ref{Feature Weights}) as it's an established fact that Random Forest is a highly interpretable classifier.
Alternatively, we run our 3-Layer Perceptron, omitting each feature in isolation, and note down the drop in classification accuracy, which should also tell us how important each feature is to determine the extremity of a group. The observations are shown in Table \ref{Accuracy_drop}.

The review count, i.e., how much a group writes about a brand, seems to be the strongest indicator of extremism. It shows that most of the times the activity involves writing a lot about a brand compared to moderate reviewers, which can be attributed to the fact that the group promotes/demotes a target brand by an overwhelming number of reviews. This trend can be explained by both the experiments: as given by random forest, and dropping this feature causes a significant drop in the 3-layer Perceptron's classification accuracy.

The rating deviation of the group seems to be another important feature for distinguishing two types of groups. Random Forest assigns it a decent score (Table \ref{Feature Weights}), and 4-Layer Perceptron also exhibits an accuracy drop on removing this feature. \textcolor{black}{This conforms to the past studies that associate extremism in review to the rating associated with the review \cite{extremity_helpful_review}.}

Also, a major difference between extreme and moderate reviews is whether the product being reviewed was actually bought by the reviewer or not, as shown by the Amazon verified purchase tag next to the review.  It shows a consistent value of importance in both our experiments.


Ratings are generally coupled with a similar sentiment in the review text itself. Therefore, it is not a very surprising observation that the average sentiment of a group to a brand also plays a minor role in the classification process, as given by the feature weight. However, since sentiment is a bit correlated to the rating, dropping this feature does not produce a great loss of accuracy since the missing information is nevertheless captured by the rating.

Unsurprisingly, average upvote count does not seem to have much impact in the nature of the group, since, for different product types, both kinds of reviews are helpful. Interestingly, `Average Rating' does not have any significant impact on the extremist and moderate review detection. This might be attributed to both kinds of groups which have a similar average over their group.

The overall trend of the extremist versus moderate reviewers is summarized in Figure \ref{fig:comparions} and explained in the following section.

\begin{figure*}
    \centering
    \includegraphics[width=0.8\linewidth]{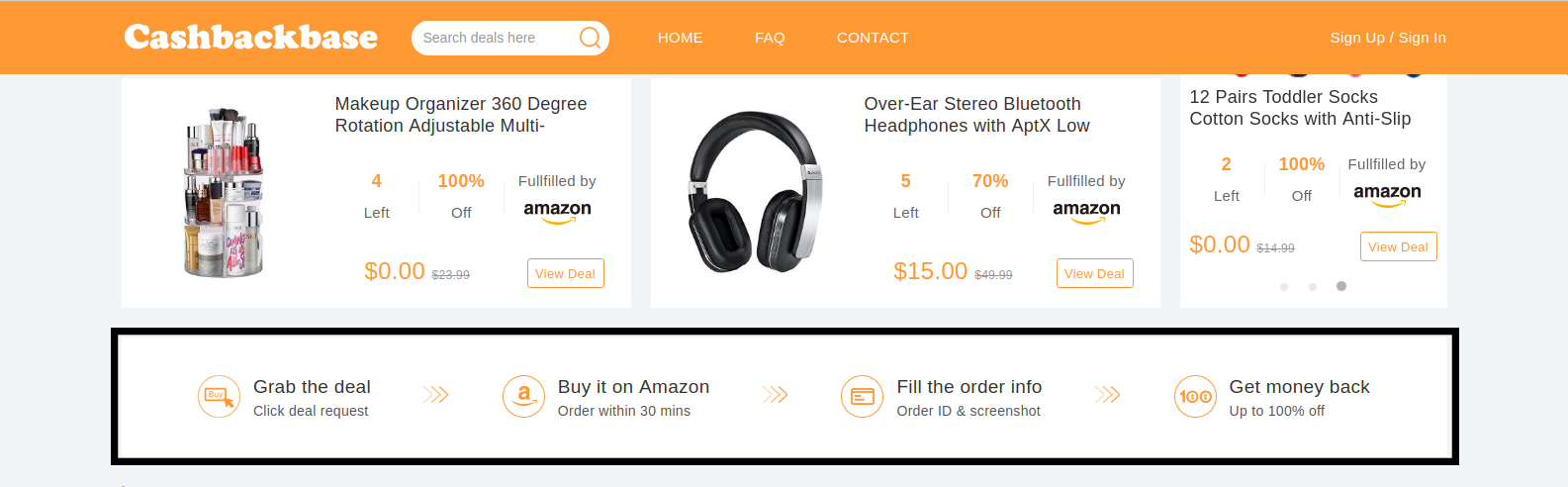}
    \caption{Dashboard for a cashback based website.}
    \label{fig:cashback}
\end{figure*}

\section{Characterization of Extremist Groups}\label{sec:characterization}
In this section, we attempt to characterize extremist reviewer groups and distinguish them from moderate groups. 
Figure \ref{fig:comparions}(a) shows that on an average, the extremist group writes a lot of 5-star reviews, while moderate users are spread around the 4-star. This is due to the nature of the problem itself, that extremist groups tend to give extreme ratings to the target brand. The lack of reviews on the extreme negative side would be discussed in later sections. 

Figure \ref{fig:comparions}(b) shows that extremist groups tend to give similar ratings to the brand, giving a very small rating deviation. On the other hand, moderate users have some variance in the given ratings. The fact that the rating deviations is about 1 and not higher may be attributed to the fact that generally people naturally tend to write positive reviews, with reviews generally not falling in the extreme negative side of the spectrum. 

Figure \ref{fig:comparions}(c) shows that most of the reviews written by extremist users are about verified purchases. This distribution is significantly skewed compared to moderate users. Apparently, this observation may seem counter-intuitive; however, this trend can be explained by the existence of several Blackmarket services which give products for free or heavily discounted rates (we will discuss it later). 

Figure \ref{fig:comparions}(d) shows that moderate users tend to write reviews using a neutral or near-neutral tone, with extremists writing reviews using a positive sentiment in their writing. 

Figure \ref{fig:comparions}(e) indicates a similar trend of group time window distribution for both types of groups. Therefore, this may not be a good feature to predict the nature of a given \emph{(group, brand)} pair.

Figure \ref{fig:comparions}(f) shows that extremist groups have a higher chance of reviewing the product of target brand earlier compared to the moderate users. 

Figure \ref{fig:comparions}(g) infers that the number of upvotes does not play a significant role, and the trend justifies the phenomenon. This may be attributed to the fact that either people do not take their time to upvote meaningful reviews, and upvotes given by the group members in case of extremist groups are nullified by upvotes given by users to moderate reviews. 

Figure \ref{fig:comparions}(h) shows that extremist groups tend to write more reviews for a target brand compared to the moderate reviewers.

\section{Discussion \label{sec:other_techniques}}

During labelling and classification, we found a fair number of spam groups showing extremist behaviour. This is a strong indicator for our hypothesis that extremism is widely prominent at the brand level, with groups aiming to promote or demote brands, due to various reasons which may include economic incentives either directly or indirectly by the brand itself.

A closer look at the target brands reveals that most of them are not widely popular with little recognition among the general public. This is not surprising since these brands may be part of start-ups that might have a constraint on resources for marketing or publicity, and thus interested in unethical practices to boost their growth. Also, since they have a relatively small consumer base, and thus a small number of moderate reviews, reviews from spam groups influence their brand image more strongly.

Well established and more reputed brands do not appear to be a target of these malicious activities due to several reasons: (1) they already have a huge customer base and followers, (2) reviews by spam groups would be suppressed by the sheer number of moderate reviews for other customers, and (3) they have lot of reputation and popularity to lose if caught indulging in malicious activities.

\textcolor{black}{Extremity in reviews is an important indicator of attitude of people towards a brand, and thus there is a need to incorporate this extremism while developing systems (such as a recommendation system or product feature summarizer) based on product reviews. This robustness can be achieved by discounting suspicious extreme and moderate reviews. One way to do so would be to rank the groups by combining the predicted extremity and collusive scores and then discount the top-most and bottom-most reviews from the group. This would help the model to focus on genuine extremist and genuinely moderate reviews and not be influenced by reviews that were incentivised with latent motives.}

There exist premium services that offer plans at different costs ranging from different reviews. Another crowd-sourcing technique is also popular on websites like \textit{Ripple Influence}\footnote{https://www.rippleinfluence.com}, or \textit{Product Elf Coupons}\footnote{http://productelf.com}, where a seller puts his/her products along with a discount coupon, which the customer (the influencer) uses to purchase the product at a discounted price and post their reviews. Doing this, they gain more reputation on these sites and then can ``unlock'' higher discount coupons. \textcolor{black}{Such kind of reviews, though effective, will not qualify as a verified-purchase review since they were received at discounted rates, and systems can easily prevent spammer groups by using only verified reviews. Therefore, spammers now require a different strategy.}

Since Amazon had restricted reviews recently in 2016, the services for obtaining reviews are forced to modify their mode of operation. A way to circumvent \textcolor{black}{the restriction imposed on the system} is to provide discount/refund off-site (e.g., a cashback). For example, we show a dashboard for a cashback based website (\url{cashbackbase.com}) in Figure \ref{fig:cashback}. At the bottom, we can see their process for issuing cashback: A reviewer needs to ``claim'' the discount on the site, purchase the product within a time limit, and upload information proving the purchase on Amazon. After verifying the purchase, the website sanctions a cashback into the buyer's account. Thus the buyer effectively was incentivised for the product purchase. Although the website does not mandate reviews from the buyers, they do claim that their websites can help in gathering reviews (on the merchant portal). On further investigation, we came to know that this is indeed the case. One such confession by an anonymized top reviewer was seen in a blog post \cite{blog}. \textcolor{black}{Thus, it is to be expected that the online marketplace will be infested with manipulated extreme review instances. It is even more difficult to distinguish such cases due to the off-site nature, but by relating brand level group extremism activities to designing robust and feasible systems becomes possible.}

Finally, we find that almost all the extremist groups are involved in promoting their target brand instead of demoting others. This observation may be attributed to the reason that it is more effective and profitable to edge out the competition by boosting their own brand image, rather than working to hinder the competition since that would require influencing a lot of brands, requiring a lot more resources and time. 






\section{Conclusion}\label{sec:conclusion}
In this paper, we discussed an unexplored form of opinion spam, where spammers target brands as a whole, posting extreme reviews, to change the overall sentiment about the brand. These groups are often part of a complex business web that is capable of influencing the overall popularity and reputation of several brands on review websites. \textcolor{black}{This study is the first step towards linking brand-level group activities and extremism in reviews, which uncovers important insights about marketplace activities. These insights would help in developing a better recommendation that make use of online reviews.} 

A set of candidate spam groups was retrieved using FIM, and extremist groups were identified by observing their actions as a group based on various features, using a supervised learning technique based on a ground-truth of manually annotated labels. We then classified extremist and moderate groups and compared the accuracy across multiple classification methods.

After classifying these groups, we observed behaviours for extremist groups in detail to gain further insights about the phenomenon and the overall trends of how these groups target these brands. We have also released the codes and annotated dataset for further studies.

\section*{Acknowledgement}
The authors would like to acknowledge the support of Ramanujan Fellowship, DST (ECR/2017/00l691) and the Infosys Centre of AI, IIIT-Delhi, India.


\begin{IEEEbiography}[{\includegraphics[width=1in,height=1.25in,clip,keepaspectratio]{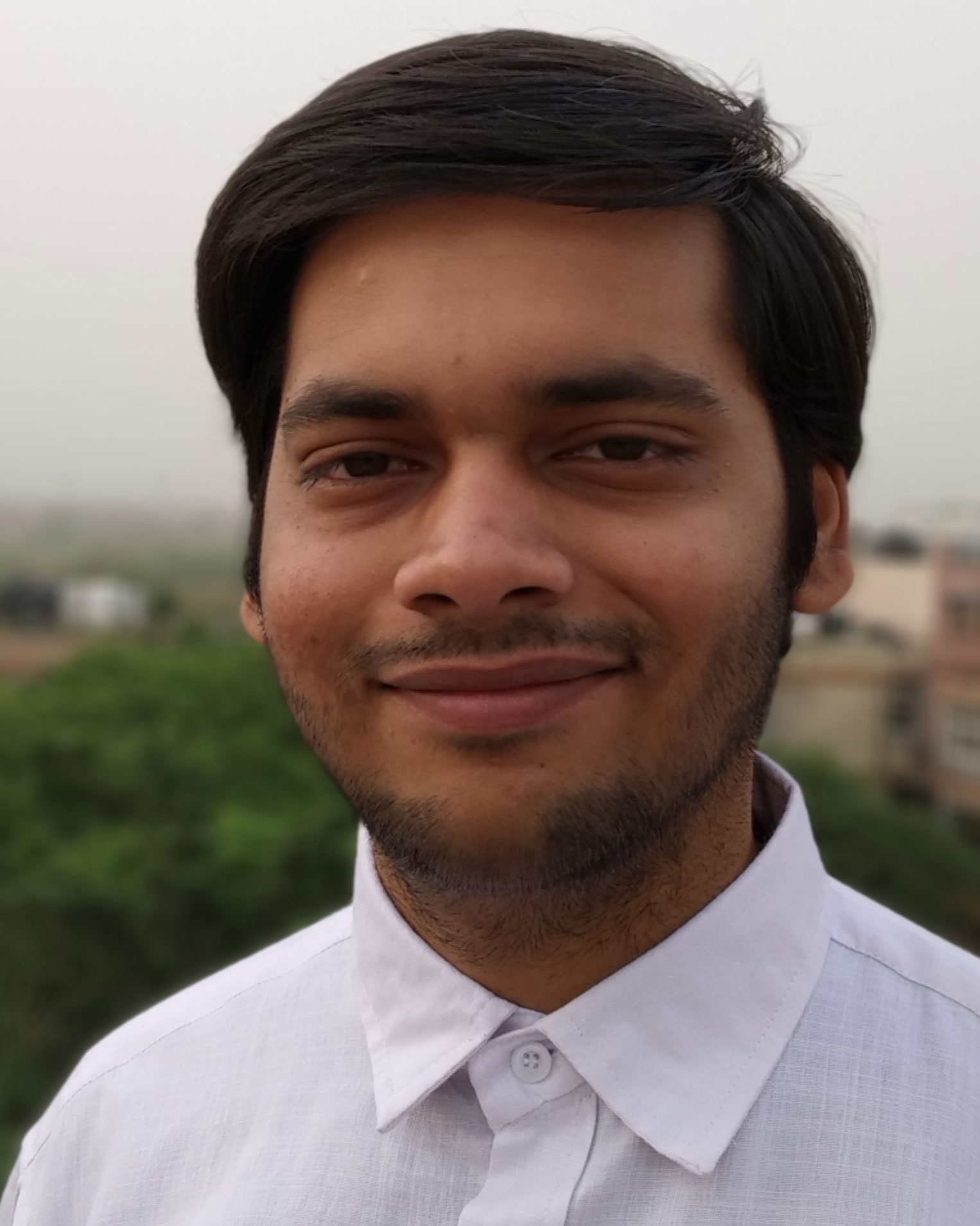}}]{Viresh Gupta} is a final year undergraduate student at the Dept. of Computer Science and Engineering, IIIT Delhi, India. His research focuses on Natural Language Processing, Complex Network Analysis, Graph Mining, Representation Learning, Multimodal Analysis and applications of data-driven Artificial Intelligence methods like Machine Learning and Deep Learning in problems across diverse domains. He has received recognition in the form of dean's list award for both academics and research and development at IIIT-Delhi.
\end{IEEEbiography}

\begin{IEEEbiography}[{\includegraphics[width=1in,height=1.25in,clip,keepaspectratio]{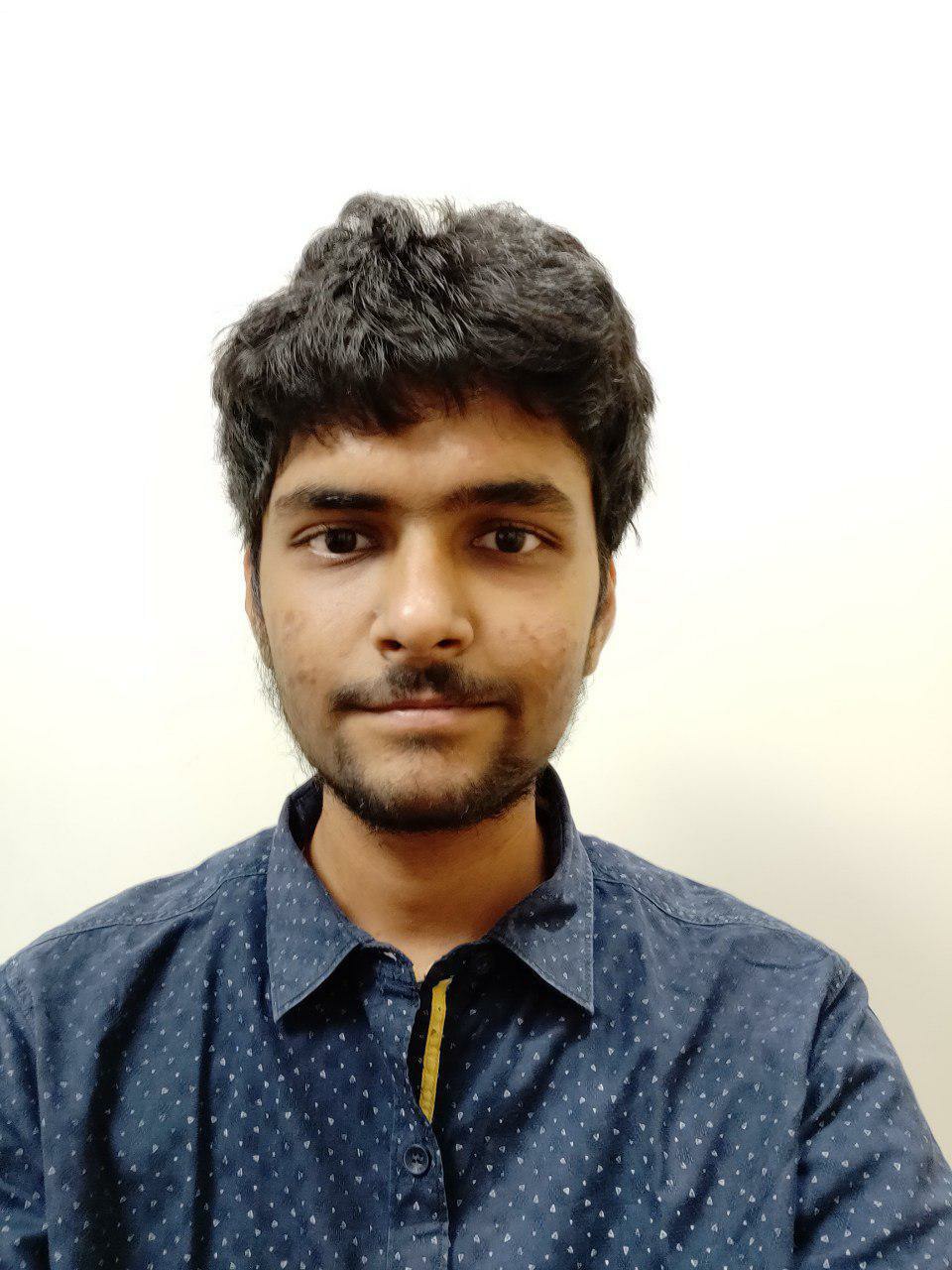}}]{Aayush Aggarwal} is a final year undergraduate student at the Dept. of Computer Science and Engineering, IIIT-Delhi, India. His research interests include Data Mining, Natural Language Processing, and Artificial Intelligence, who is also exploring domains related to theoretical computer science and mathematics.
\end{IEEEbiography}

\begin{IEEEbiography}[{\includegraphics[width=1in,height=1.25in,clip,keepaspectratio]{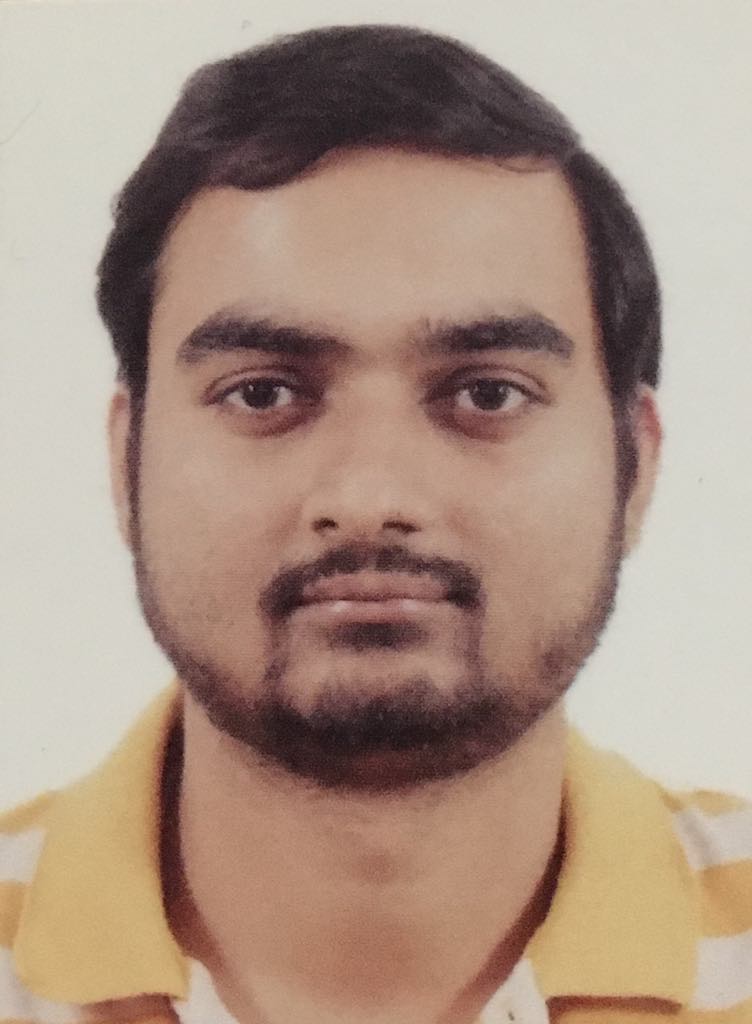}}]{Tanmoy Chakraborty} is an Assistant Professor and a Ramanujan Fellow at the Dept. of Computer Science and Engineering, IIIT-Delhi, India. His primary research interests include Social Network Analysis, Data Mining, and Natural Language Processing. He has received several awards including Google Indian Faculty Award, Early Career Research Award, DAAD Faculty award. 
\end{IEEEbiography}


\begin{thebibliography}{10}
	\providecommand{\url}[1]{#1}
	\csname url@samestyle\endcsname
	\providecommand{\newblock}{\relax}
	\providecommand{\bibinfo}[2]{#2}
	\providecommand{\BIBentrySTDinterwordspacing}{\spaceskip=0pt\relax}
	\providecommand{\BIBentryALTinterwordstretchfactor}{4}
	\providecommand{\BIBentryALTinterwordspacing}{\spaceskip=\fontdimen2\font plus
		\BIBentryALTinterwordstretchfactor\fontdimen3\font minus
		\fontdimen4\font\relax}
	\providecommand{\BIBforeignlanguage}[2]{{%
			\expandafter\ifx\csname l@#1\endcsname\relax
			\typeout{** WARNING: IEEEtran.bst: No hyphenation pattern has been}%
			\typeout{** loaded for the language `#1'. Using the pattern for}%
			\typeout{** the default language instead.}%
			\else
			\language=\csname l@#1\endcsname
			\fi
			#2}}
	\providecommand{\BIBdecl}{\relax}
	\BIBdecl
	
	\bibitem{quoteLink}
	A.~Kim, ``{That review you wrote on Amazon? Priceless},''
	\url{https://www.usatoday.com/story/tech/news/2017/03/20/review-you-wrote-amazon-pric
		ess/99332602/}, 2017.
	
	\bibitem{review_echo_chamber}
	\BIBentryALTinterwordspacing
	E.~Gilbert and K.~Karahalios, ``Understanding deja reviewers,'' in \emph{CSCW},
	2010, pp. 225--228. [Online]. Available:
	\url{http://doi.acm.org/10.1145/1718918.1718961}
	\BIBentrySTDinterwordspacing
	
	\bibitem{ReviewLimit}
	Amazon.in, ``{Review Community Guidelines},''
	\url{https://www.amazon.in/gp/help/customer/display.html?nodeId=201929730},
	2018, online.
	
	\bibitem{GSRank}
	A.~Mukherjee, B.~Liu, and N.~Glance, ``Spotting fake reviewer groups in
	consumer reviews,'' in \emph{WWW}, 2012, pp. 191--200.
	
	\bibitem{Lu_annotation_simult}
	Y.~Lu, L.~Zhang, Y.~Xiao, and Y.~Li, ``Simultaneously detecting fake reviews
	and review spammers using factor graph model,'' in \emph{WebSci}, 2013, pp.
	225--233.
	
	\bibitem{Speagle}
	S.~Rayana and L.~Akoglu, ``Collective opinion spam detection: Bridging review
	networks and metadata,'' in \emph{KDD}, 2015, pp. 985--994.
	
	\bibitem{sarthikaijcai}
	\BIBentryALTinterwordspacing
	S.~Dhawan, S.~C.~R. Gangireddy, S.~Kumar, and T.~Chakraborty, ``Spotting
	collusive behaviour of online fraud groups in customer reviews,''
	\emph{CoRR}, vol. abs/1905.13649, 2019. [Online]. Available:
	\url{http://arxiv.org/abs/1905.13649}
	\BIBentrySTDinterwordspacing
	
	\bibitem{Dave:2003:MPG:775152.775226}
	K.~Dave, S.~Lawrence, and D.~M. Pennock, ``Mining the peanut gallery: Opinion
	extraction and semantic classification of product reviews,'' in \emph{WWW},
	2003, pp. 519--528.
	
	\bibitem{Pang:2002:TUS:1118693.1118704}
	B.~Pang, L.~Lee, and S.~Vaithyanathan, ``Thumbs up?: Sentiment classification
	using machine learning techniques,'' in \emph{EMNLP}, 2002, pp. 79--86.
	
	\bibitem{6508366}
	K.~{Mouthami}, K.~N. {Devi}, and V.~M. {Bhaskaran}, ``Sentiment analysis and
	classification based on textual reviews,'' in \emph{ICICES}, 2013, pp.
	271--276.
	
	\bibitem{YE20096527}
	Q.~Ye, Z.~Zhang, and R.~Law, ``Sentiment classification of online reviews to
	travel destinations by supervised machine learning approaches,'' \emph{Expert
		Systems with Applications}, vol.~36, no. 3, Part 2, pp. 6527 -- 6535, 2009.
	
	\bibitem{Chelliah:2017:PRE:3109859.3109936}
	M.~Chelliah and S.~Sarkar, ``Product recommendations enhanced with reviews,''
	in \emph{RecSys}, 2017, pp. 398--399.
	
	\bibitem{CHEN201344}
	L.~Chen and F.~Wang, ``Preference-based clustering reviews for augmenting
	e-commerce recommendation,'' \emph{Knowledge-Based Systems}, vol.~50, pp. 44
	-- 59, 2013.
	
	\bibitem{recommendation_with_review}
	J.~Feuerbach, B.~Loepp, C.-M. Barbu, and J.~Ziegler, ``Enhancing an interactive
	recommendation system with review-based information filtering,'' in
	\emph{IntRS@RecSys}, 2017.
	
	\bibitem{review_with_collabfiltering}
	\BIBentryALTinterwordspacing
	A.~Almahairi, K.~Kastner, K.~Cho, and A.~C. Courville, ``Learning distributed
	representations from reviews for collaborative filtering,'' \emph{CoRR}, vol.
	abs/1806.06875, 2018. [Online]. Available:
	\url{http://arxiv.org/abs/1806.06875}
	\BIBentrySTDinterwordspacing
	
	\bibitem{Popescu:2005:EPF:1220575.1220618}
	A.-M. Popescu and O.~Etzioni, ``Extracting product features and opinions from
	reviews,'' in \emph{HLT}, 2005, pp. 339--346.
	
	\bibitem{Liu:2005:OOA:1060745.1060797}
	B.~Liu, M.~Hu, and J.~Cheng, ``Opinion observer: Analyzing and comparing
	opinions on the web,'' in \emph{WWW}, 2005, pp. 342--351.
	
	\bibitem{Hu:2004:MOF:1597148.1597269}
	M.~Hu and B.~Liu, ``Mining opinion features in customer reviews,'' in
	\emph{AAAI}, 2004, pp. 755--760.
	
	\bibitem{explain_recommendation}
	\BIBentryALTinterwordspacing
	T.~Donkers, B.~Loepp, and J.~Ziegler, ``Explaining recommendations by means of
	user reviews,'' in \emph{Proceedings of the 1st Workshop on Explainable Smart
		Systems (ExSS)}, 2018. [Online]. Available:
	\url{http://ceur-ws.org/Vol-2068/exss8.pdf}
	\BIBentrySTDinterwordspacing
	
	\bibitem{OpinionMiningAndSentimentAnalysis}
	B.~Pang, L.~Lee \emph{et~al.}, ``Opinion mining and sentiment analysis,''
	\emph{Foundations and Trends{\textregistered} in Information Retrieval},
	vol.~2, no. 1--2, pp. 1--135, 2008.
	
	\bibitem{Hu:2004:MSC:1014052.1014073}
	M.~Hu and B.~Liu, ``Mining and summarizing customer reviews,'' in \emph{KDD},
	2004, pp. 168--177.
	
	\bibitem{Somprasertsri2010MiningFI}
	G.~Somprasertsri and P.~Lalitrojwong, ``Mining feature-opinion in online
	customer reviews for opinion summarization,'' \emph{J. UCS}, vol.~16, pp.
	938--955, 2010.
	
	\bibitem{Zhuang:2006:MRM:1183614.1183625}
	L.~Zhuang, F.~Jing, and X.-Y. Zhu, ``Movie review mining and summarization,''
	in \emph{CIKM}, 2006, pp. 43--50.
	
	\bibitem{pecar-2018-towards}
	\BIBentryALTinterwordspacing
	S.~Pecar, ``Towards opinion summarization of customer reviews,'' in
	\emph{Proceedings of {ACL}, Student Research Workshop}.\hskip 1em plus 0.5em
	minus 0.4em\relax Melbourne, Australia: Association for Computational
	Linguistics, Jul. 2018, pp. 1--8. [Online]. Available:
	\url{https://www.aclweb.org/anthology/P18-3001}
	\BIBentrySTDinterwordspacing
	
	\bibitem{HoDac2013TheEO}
	N.~N. Ho-Dac, S.~J. Carson, and W.~L. Moore, ``The effects of positive and
	negative online customer reviews : Do brand strength and category maturity
	matter ?'' 2013.
	
	\bibitem{boxofficeMarketingStudyWordOfMouth}
	P.~K. Chintagunta, S.~Gopinath, and S.~Venkataraman, ``The effects of online
	user reviews on movie box office performance: Accounting for sequential
	rollout and aggregation across local markets,'' \emph{Marketing Science},
	no.~5, pp. 944--957, 2010.
	
	\bibitem{extremity_helpful_review}
	S.~Mudambi and D.~Schuff, ``What makes a helpful online review? a study of
	customer reviews on amazon.com.'' \emph{MIS Quarterly}, vol.~34, pp.
	185--200, 03 2010.
	
	\bibitem{Yin2014ExploringHC}
	G.~Yin, L.~Wei, W.~Xu, and M.~Chen, ``Exploring heuristic cues for consumer
	perceptions of online reviews helpfulness: the case of yelp.com,'' in
	\emph{PACIS}, 2014.
	
	\bibitem{Kim:2006:AAR:1610075.1610135}
	\BIBentryALTinterwordspacing
	S.-M. Kim, P.~Pantel, T.~Chklovski, and M.~Pennacchiotti, ``Automatically
	assessing review helpfulness,'' in \emph{Proceedings of the Conference on
		Empirical Methods in Natural Language Processing}, ser. EMNLP, 2006, pp.
	423--430. [Online]. Available:
	\url{http://dl.acm.org/citation.cfm?id=1610075.1610135}
	\BIBentrySTDinterwordspacing
	
	\bibitem{Wang2019}
	\BIBentryALTinterwordspacing
	Y.~Wang, J.~Wang, and T.~Yao, ``What makes a helpful online review? a
	meta-analysis of review characteristics,'' \emph{Electronic Commerce
		Research}, vol.~19, no.~2, pp. 257--284, Jun 2019. [Online]. Available:
	\url{https://doi.org/10.1007/s10660-018-9310-2}
	\BIBentrySTDinterwordspacing
	
	\bibitem{recommendReviews}
	M.~Salehan, M.~Mousavizadeh, and M.~Koohikamali, ``A recommender system for
	online consumer reviews research-in-progress,'' 2015.
	
	\bibitem{twosided}
	M.~Eisend, ``Two-sided advertising: A meta-analysis,'' 2006.
	
	\bibitem{extreme_ebay}
	P.~A. Pavlou and A.~Dimoka, ``The nature and role of feedback text comments in
	online marketplaces: Implications for trust building, price premiums, and
	seller differentiation,'' \emph{Information Systems Research}, vol.~17,
	no.~4, pp. 392--414, dec 2006.
	
	\bibitem{extreme_book}
	C.~Forman, A.~Ghose, and B.~Wiesenfeld, ``Examining the relationship between
	reviews and sales: The role of reviewer identity disclosure in electronic
	markets,'' \emph{Information Systems Research}, vol.~19, no.~3, pp. 291--313,
	sep 2008.
	
	\bibitem{Krosnick1993}
	\BIBentryALTinterwordspacing
	J.~A. Krosnick, D.~S. Boninger, Y.~C. Chuang, M.~K. Berent, and et~al,
	``Attitude strength: One construct or many related constructs?''
	\emph{Journal of Personality and Social Psychology}, vol.~65, no.~6, pp.
	1132--1151, 1993. [Online]. Available:
	\url{https://doi.org/10.1037/0022-3514.65.6.1132}
	\BIBentrySTDinterwordspacing
	
	\bibitem{Kaplan_1972}
	K.~J. Kaplan, ``On the ambivalence-indifference problem in attitude theory and
	measurement: A suggested modification of the semantic differential
	technique.'' \emph{Psychological Bulletin}, vol.~77, no.~5, pp. 361--372,
	1972.
	
	\bibitem{Presser_1980}
	\BIBentryALTinterwordspacing
	S.~Presser and H.~Schuman, ``The measurement of a middle position in attitude
	surveys,'' \emph{Public Opinion Quarterly}, vol.~44, no.~1, p.~70, 1980.
	[Online]. Available: \url{https://doi.org/10.1086%2F268567}
		\BIBentrySTDinterwordspacing
		
		\bibitem{WSDMJindal}
		N.~Jindal and B.~Liu, ``Opinion spam and analysis,'' in \emph{WSDM}, 2008, pp.
		219--230.
		
		\bibitem{Ottreviews2}
		M.~Ott, Y.~Choi, C.~Cardie, and J.~T. Hancock, ``Finding deceptive opinion spam
		by any stretch of the imagination,'' in \emph{ACL}, 2011, pp. 309--319.
		
		\bibitem{onlyReviewSpam2}
		N.~Jindal, B.~Liu, and E.-P. Lim, ``Finding unusual review patterns using
		unexpected rules,'' in \emph{CIKM}, 2010, pp. 1549--1552.
		
		\bibitem{onlyReviewSpam3}
		F.~Li, M.~Huang, Y.~Yang, and X.~Zhu, ``Learning to identify review spam,'' in
		\emph{IJCAI}, 2011, pp. 2488--2493.
		
		\bibitem{MarkovFields}
		A.~Fayazi, K.~Lee, J.~Caverlee, and A.~Squicciarini, ``Uncovering crowdsourced
		manipulation of online reviews,'' in \emph{SIGIR}, 2015, pp. 233--242.
		
		\bibitem{ratingBehaviours}
		E.-P. Lim, V.-A. Nguyen, N.~Jindal, B.~Liu, and H.~W. Lauw, ``Detecting product
		review spammers using rating behaviors,'' in \emph{CIKM}, 2010, pp. 939--948.
		
		\bibitem{onlyReviewer1}
		G.~Wang, S.~Xie, B.~Liu, and P.~S. Yu, ``Review graph based online store review
		spammer detection,'' in \emph{ICDM}, 2011, pp. 1242--1247.
		
		\bibitem{onlyReviewer2}
		A.~Mukherjee, V.~V. Venkataraman, B.~Liu, and N.~S. Glance, ``What yelp fake
		review filter might be doing?'' in \emph{ICWSM}, 2013, pp. 409--418.
		
		\bibitem{MukherjeeFootprinting}
		A.~Mukherjee, A.~Kumar, B.~Liu, J.~Wang, M.~Hsu, M.~Castellanos, and R.~Ghosh,
		``Spotting opinion spammers using behavioral footprints,'' in \emph{KDD},
		2013, pp. 632--640.
		
		\bibitem{burstiness}
		G.~Fei, A.~Mukherjee, B.~Liu, M.~Hsu, M.~Castellanos, and R.~Ghosh,
		``Exploiting burstiness in reviews for review spammer detection,''
		\emph{ICWSM}, pp. 175--184, 01 2013.
		
		\bibitem{kumar2018rev2}
		S.~Kumar, B.~Hooi, D.~Makhija, M.~Kumar, C.~Faloutsos, and V.~Subrahmanian,
		``Rev2: Fraudulent user prediction in rating platforms,'' in
		\emph{Proceedings of the Eleventh ACM International Conference on Web Search
			and Data Mining}.\hskip 1em plus 0.5em minus 0.4em\relax ACM, 2018, pp.
		333--341.
		
		\bibitem{buyaccountbatches}
		A.~Molavi~Kakhki, C.~Kliman-Silver, and A.~Mislove, ``Iolaus: Securing online
		content rating systems,'' in \emph{WWW}, 2013, pp. 919--930.
		
		\bibitem{Xuhcbm}
		C.~Xu and J.~Zhang, ``Towards collusive fraud detection in online reviews,'' in
		\emph{ICDM}, 2015, pp. 1051--1056.
		
		\bibitem{GGSpam}
		Z.~Wang, S.~Gu, X.~Zhao, and X.~Xu, ``Graph-based review spammer group
		detection,'' \emph{Knowl. Inf. Syst.}, pp. 571--597, 2018.
		
		\bibitem{relim}
		C.~Borgelt, ``Frequent item set mining,'' \emph{Wiley Int. Rev. Data Min. and
			Knowl. Disc.}, pp. 437--456, 2012.
		
		\bibitem{spmf}
		P.~Fournier-Viger, J.~C.-W. Lin, A.~Gomariz, T.~Gueniche, A.~Soltani, Z.~Deng,
		and H.~T. Lam, ``The spmf open-source data mining library version 2,'' in
		\emph{Machine Learning and Knowledge Discovery in Databases}, 2016, pp.
		36--40.
		
		\bibitem{clustering}
		C.~Xu, ``Detecting collusive spammers in online review communities,'' in
		\emph{Proceedings of the sixth workshop on Ph. D. students in information and
			knowledge management}.\hskip 1em plus 0.5em minus 0.4em\relax ACM, 2013, pp.
		33--40.
		
		\bibitem{liang_annotation1}
		D.~{Liang}, X.~{Liu}, and H.~{Shen}, ``Detecting spam reviewers by combing
		reviewer feature and relationship,'' in \emph{ICCSS}, 2014, pp. 102--107.
		
		\bibitem{Xie_annotation}
		S.~Xie, G.~Wang, S.~Lin, and P.~S. Yu, ``Review spam detection via temporal
		pattern discovery,'' in \emph{KDD}, 2012, pp. 823--831.
		
		\bibitem{annotation}
		A.-T. Pieper, ``Detecting review spam on amazon with reviewalarm,'' {B.S.}
		thesis, University of Twente, 2016.
		
		\bibitem{mukherjee2012spotting}
		A.~Mukherjee, B.~Liu, and N.~Glance, ``Spotting fake reviewer groups in
		consumer reviews,'' in \emph{Proceedings of the 21st international conference
			on World Wide Web}.\hskip 1em plus 0.5em minus 0.4em\relax ACM, 2012, pp.
		191--200.
		
		\bibitem{fraudeagle}
		L.~Akoglu, R.~Chandy, and C.~Faloutsos, ``Opinion fraud detection in online
		reviews by network effects,'' in \emph{ICWSM}, 2013.
		
		\bibitem{sentiwordnet3}
		S.~Baccianella, A.~Esuli, and F.~Sebastiani, ``Sentiwordnet 3.0: An enhanced
		lexical resource for sentiment analysis and opinion mining,'' in \emph{LREC},
		2010.
		
		\bibitem{8793010}
		H.~{Fulara}, G.~{Singh}, D.~{Jaisinghani}, M.~{Maity}, T.~{Chakraborty}, and
		V.~{Naik}, ``Use of machine learning to detect causes of unnecessary active
		scanning in wifi networks,'' in \emph{WoWMoM}, 2019, pp. 1--9.
		
		\bibitem{blog}
		Y.~Chen, ``Confessions of a paid amazon review writer,''
		https://digiday.com/marketing/vendors-ask-go-around-policy-confessions-top-ranked-amazon-review-writer/.
		
	\end{thebibliography}
\end{document}